\documentclass{aa}

\usepackage{fixltx2e}
\usepackage{graphicx}
\usepackage{txfonts}
\usepackage{rotating}
 
\newcommand{\de}{{\rm d}}

\begin{document}

\title{Methanol masers reveal the magnetic field \\of the high-mass protostar IRAS\,18089-1732}
\titlerunning{Magnetic field in IRAS\,18089}
\author{D. Dall'Olio\inst{1}
          \and
          W. H. T. Vlemmings\inst{1}
          \and
          G. Surcis\inst{2}
          \and
          H. Beuther\inst{3}
          \and
          B. Lankhaar\inst{1}
          \and
          M. V. Persson\inst{1}
          \and
          A. M. S. Richards\inst{4}
          \and
          E. Varenius\inst{1}
          }

          \institute{Department of Space, Earth and Environment, Chalmers University of Technology,
            Onsala Space Observatory,\\ Observatoriev\"agen 90, 43992 Onsala, Sweden;
             \email{daria.dallolio@chalmers.se}
             \and
             INAF--Osservatorio Astronomico di Cagliari, Via della Scienza 5, 09047 Selargius, Italy
             \and
             Max-Planck-Institute for Astronomy, K{\"o}nigstuhl 17, 69117 Heidelberg, Germany
             \and
             Jodrell Bank Centre for Astrophysics, Department of Physics and Astronomy, University of Manchester,  M139PL Manchester, UK}
   \date{Received 2 June 2017; accepted 9 August 2017}
  \abstract
  {The importance of the magnetic field in high-mass-star formation is
    not yet fully clear and there are still many open questions
    concerning its role in the accretion processes and generation of
    jets and outflows. In the past few years, masers have been
    successfully used to probe the magnetic field morphology and
    strength at scales of a few au around massive protostars, by
    measuring linear polarisation angles and Zeeman splitting.  The
    massive protostar IRAS\,18089-1732 is a well studied high-mass-star forming region, showing a hot core chemistry and a
    disc-outflow system. Previous SMA observations of polarised dust
    revealed an ordered magnetic field oriented around the disc of
    IRAS\,18089-1732.}
  {We want to determine the magnetic field in the dense region probed
    by 6.7 GHz methanol maser observations and compare it with
    observations in dust continuum polarisation, to investigate how
    the magnetic field in the compact maser region relates to the
    large-scale field around massive protostars.}
  {We reduced MERLIN observations at 6.7 GHz of IRAS\,18089-1732 and
    we analysed the polarised emission by methanol masers.}
   {Our MERLIN observations show that the magnetic field in the 6.7
     GHz methanol maser region is consistent with the magnetic field
     constrained by the SMA dust polarisation observations. A
     tentative detection of circularly polarised line emission is also
     presented.}
   {We found that the magnetic field in the maser region has the same
     orientation as in the disk. Thus the large-scale field component,
     even at the au scale of the masers, dominates over any small-scale field fluctuations. We obtained, from the circular
     polarisation tentative detection, a field strength along the line
     of sight of 5.5 mG which appeared to be consistent with the
     previous estimates.}
   \keywords{magnetic field --
                stars: formation -- stars: massive --
                masers -- polarization}

   \maketitle

\section{Introduction}

The role of magnetic fields during the formation of high-mass stars is
not yet fully understood.  As in the case of low-mass star formation,
simulations have shown that the magnetic field appears to prevent
fragmentation around massive protostars (\citealt{Peters2011},
\citealt{Myers2013}) to influence accretion and to drive feedback
phenomena such as collimated outflows and jets
(\citealt{Seifried2011,Seifried2012}).  Both the core accretion model
(e.g.\ \citealt{McKeeTan2002}, \citealt{Banerjee2007}) and the
competitive accretion model (e.g.\ \citealt{Bonnell2006}) need
observational constraints on magnetic fields to properly investigate
their effect on the high-mass star formation process \citep{Tan2014}.

Several fine-tuned models have shown that some detailed and specific
magneto-hydrodynamic (MHD) configurations, yet to be tested
observationally, can explain observed
morphologies (e.g. \citealt{Krumholz2013}; \citealt{Li2014};
\citealt{Seifried2015} and references therein).  A typical example is
a rotating Keplerian disc around a protostar. While several
circumstellar discs have been observed around massive protostars
(\citealt{Cesaroni2006,Cesaroni2007,Beltran2016}),
theoretical studies have found it
difficult to form such discs due to strong magnetic braking which
removes most of the angular momentum from the circumstellar
gas. This is the so-called magnetic braking catastrophe in disc
formation (\citealt{Mestel-Spitzer1956}; \citealt{Allen2003};
\citealt{Mellon-Li2008}; \citealt{Li2011, Li2013}). Only by inserting non-ideal
MHD effects such as Ohmic dissipation or ambipolar diffusion, or the
combined action of both, is it possible to overcome the
apparent conflict between observations and simulations
(\citealt{Machida2014}, \citealt{Zhao2016}). Moreover, \citet{Zhao2016} have
shown that chemistry and microscopic physical processes, including
advection of gas phase and grain species as well as grain evolution,
must also be inserted in non-ideal MHD simulations to obtain a more
realistic picture of the behaviour of a strongly magnetised
core. However, it is not yet clear which process dominates between Ohmic
dissipation and ambipolar diffusion; it probably depends on the
magnitude of the initial magnetic energy density relative to the
gravitational and turbulent energy density and the initial magnetic
field configuration. Thus, we need to probe the magnetic field at
small scales.
The identification of high-mass protostars is however extremely
complicated due to their fast evolution, and their 
location inside distant, dense, and dark clusters.

The first full polarisation observations of 6.7 GHz methanol masers
were made by \citet{Ellingsen2002}. Thereafter, \citet{Green2007}
\citet{Vlemmings2006, Vlemmings2008, Vlemmings2010}, \citet{Dodson2012} and
\citet{Surcis2012, Surcis2014iras20126, Surcis2015} have demonstrated that
maser emission (from e.g. methanol and water) allows us to probe
magnetic fields. Through the study of linear and circular polarised
emission, it is possible to obtain the strength, morphology, and
evolution of the magnetic field on scale comparable to circumstellar
discs ($\sim100$ au).  However, this has been done only in a limited
number of cases which still prevents building a complete picture of
the role played by magnetic fields. Most importantly, what is still
lacking is more observational evidence that the magnetic field at
small scales probed by masers represents the field at larger scales,
probed, for example,\ by the dust, and not small-scale fluctuations. Currently,
few observations of both masers and dust polarisation exist
towards the same regions (e.g. \citealt{Surcis2014iras20126}).

The high-mass protostar IRAS\,18089-1732 is particularly important
because observations of dust emission have already shown the structure
of the magnetic field at large scales ($\sim5000$ au,
\citealt{Beuther2010}). In this paper, we investigate its small-scale
magnetic field, by analysing a three-epoch Multi-Element Radio Linked
Interferometer Network (MERLIN) observation of the 6.7~GHz CH$_3$OH
(methanol) maser generated in the same region of the disc.  We use
large scale to refer to arcsec scales within a $\sim4\arcsec$ region
centred on IRAS\,18089-1732, and small scale for scales for a typical
maser region size of $\sim10$ mas within the same region. We present the
first polarised map of the masers for IRAS\,18089-1732 and we show
that the small-scale magnetic field probed by the masers is consistent
with the large-scale magnetic field traced by the dust.

The structure of this paper is as follows.  We introduce
IRAS\,18089-1732 in Sect.~\ref{sec:iras}.  We describe the
observations and data analysis in Sect.~\ref{sec:observ-data-red}. We
present our result in Sect.~\ref{sec:results} and discuss them in
Sect.~\ref{sec:discussion}. In Sect.~\ref{sec:conclusions} we give
our conclusions and future perspectives.

\section{The case of IRAS\,18089-1732}
\label{sec:iras}
\object{IRAS\,18089-1732} (hereafter IRAS\,18089) is a well studied
protostar presenting a velocity v$_{lsr}$=33.8 km s$^{-1}$
\citep{Beuther2005} and located at a distance of 2.34 kpc
\citep{Xu2011}. It has a luminosity $L=1.3\times10^{4}~L_\odot$
(\citealt{Sridharan2002}; rescaled to the adopted distance) and a
gaseous mass $M\sim1000~M_{\odot}$ estimated from single-dish
millimetre continuum observations (\citealt{Beuther2002}; also
rescaled to the adopted distance).

IRAS\,18089 presents the typical chemistry of `hot cores' with a line
forest profile and strong molecular emission, coming from, for example, HCOOCH$_3$, H$_2$S, SO, and SO$_2$ \citep{Beuther2004,
  Isokoski2013}.  The source is also a well-known disc-outflow
system. Submillimiter Array (SMA) observed a SiO(5-4) molecular
outflow in an approximately north-south direction \citep{Beuther2004b}
and showed rotational signatures in many molecular lines typical of an
accreting disc in the dense gas perpendicular to the outflow
\citep{Beuther2005,Zapata2006,BeutherWalsh2008}. Moreover, the
Goldreich-Kylafis effect \citep{Goldreich1981,Goldreich1982} was
detected for the CO(3-2) transition by \citet{Beuther2010}, revealing
a linear polarisation fraction up to 8\%. Furthermore,
\citet{Beuther2010} showed that the magnetic field structure is
largely aligned with the jet-outflow orientation, from the smaller
scales of the core to the larger scales of the outflow. In addition,
\citet{Beuther2010} estimated a magnetic field strength in the plane
of the sky of $B_{pos} \sim 11$~mG at a core density of
$5\times10^7$~cm$^{-3}$. This value was estimated from the analysis of
the polarised dust continuum emission observed with the SMA at 880
$\mu$m.

\citet{Vlemmings2008} derived a comparable line-of-sight magnetic
field strength $B_{los} \sim 8$~mG from the Zeeman splitting of the
6.7 GHz CH$_3$OH maser line, at densities $>10^6$~cm$^{-3}$.
Therefore \citet{Beuther2010} finally estimated a total magnetic field
strength $B_{tot}\sim \sqrt{B_{pos}^2+B_{los}^2} \sim 14$~mG, which is of the
same order of measurements made by \citet{Girart2009},
\citet{Surcis2009}, and \citet{Vlemmings2010} for similar
sources. However the estimation by \citet{Beuther2010} was made using an extrapolation of
the g-factor, obtained from measurements of 25 GHz methanol
transitions. This may result in a derived magnetic field strength that
is larger by an order of magnitude with respect to the true field
strength, as described by \citet{Vlemmings2011}.

\citet{Walsh1998} provided a map of IRAS\,18089 masers from Australia
Telescope Compact Array (ATCA) observations and obtained relative and
absolute positions with an accuracy of around $0.05\arcsec$ and
$1\arcsec$ respectively. \citet{Goedhart2009}, monitoring the
variability in IRAS\,18089, reported a periodicity of the flares
maxima of around 29.5$\pm$0.1 days, derived after 9 years of
observations with Hartebeesthoek Radio Astronomy Observatory 26 m
telescope.

\section{Observations and data reduction}
\label{sec:observ-data-red}

IRAS\,18089-1732 was observed by MERLIN at 6.7 GHz, in March, April,
and July 2008, and the data were stored in three datasets. The
observations were obtained with a single spectral window with 255
channels, covering a bandwidth of 249 kHz in March and April, and a
bandwidth of 498 kHz in July.
The total on-source observing time was 28 hours,
7 hours in March and July, and 14 hours in April. Six antennas were
used for the first two observations, and five antennas for the last
observation when Defford was not included. The longest baseline of
MERLIN is 217 km. The observational details are reported in
Table~\ref{tab:observational-details}.

\begin{table*}
  \caption[]{Observational details for IRAS\,18089-1732.}
  \label{tab:observational-details}
  \centering
  \begin{tabular}{cccccccc}
    \hline\hline
    Observation     & Polarisation   & Bandwidth\tablefootmark{a} & Channel               & Beam size                & \!\!\!Position Angle\!\!\! & RMS\tablefootmark{b}                  \\
        date        & mode           & (kHz)         & spacing (km s$^{-1}$) & (arcsec $\times$ arcsec) & ($\circ$)                  & (Jy beam$^{-1}$)     \\
    \hline
    13 March 2008   & RR, LL, RL, LR & 249           & $\sim$0.05            & 0.18 $\times$ 0.03       & 10.11                      & 0.03  \\
    7--8 April 2008 & RR, LL, RL, LR & 249           & $\sim$0.05            & 0.18 $\times$ 0.03       & 12.39                      & 0.02 \\
    4 July 2008     & RR, LL         & 498           & $\sim$0.09            & 0.24 $\times$ 0.08       & -16.34                     & 0.04 \\
    \hline
    \end{tabular}
    \tablefoot{
      \tablefoottext{a}{The spectral window included 255 channels.}
      \tablefoottext{b}{RMS on the line-free channels.}
  }
\end{table*}

\begin{table}
  \caption[]{Fluxes in Jy of the calibrators used in the three epochs.}
  \label{tab:calfluxes}
  \centering
  \begin{tabular}{cccc}
    \hline\hline
    Observation     & MRC 1757-150 & 3C84       & 3C286            \\
        date        & (observed)   & (observed) & (model) \\
    \hline
    13 March 2008   & 0.14         & 10.32      & 5.70             \\
    7--8 April 2008 & 0.14         & 14.26      & 5.70             \\
    4 July 2008     & 0.15         & 15.14      & 5.70             \\
    \hline
  \end{tabular}
\end{table}
The datasets were reduced using the Astronomical Image Processing Software
(AIPS version DEC 2016), and the
calibration was performed using 3C286, MRC 1757-150 and 3C84. 3C286
was chosen as flux and polarisation angle calibrator, MRC 1757-150 as the phase
calibrator, and 3C84 was used to calibrate
the bandpass. Since 3C286 is known to be resolved,
for the flux calibration we used a model of 3C286
provided by the MERLIN
database\footnote{http://www.e-merlin.ac.uk/data-red/}.

The March and July observations each consisted of a single run, while
two runs were performed in April, over two consecutive
days. In March and July the phase calibrator and the target
  were observed alternately for respectively 2 and 6 minutes, while in April for
  respectively 1 and 8 minutes, over a total time of about 7 hours per run.  In the
second run the phase calibration failed due to an unusual
observational strategy, so we used a model based on the strongest maser of
the first run to self calibrate the second run, using a
  solution interval of 10 minutes. For an overview of the calibration
strategy we refer to the MERLIN User
Guide\footnote{http://www.e-merlin.ac.uk/user-guide/}. In Table
\ref{tab:calfluxes} we report the fluxes of each calibrator.  The
Local Standard of Rest (LSR) correction was applied to ensure constant velocity in the
target frame.

\begin{figure*}[htb]
   \centering
   \includegraphics[width=\textwidth]{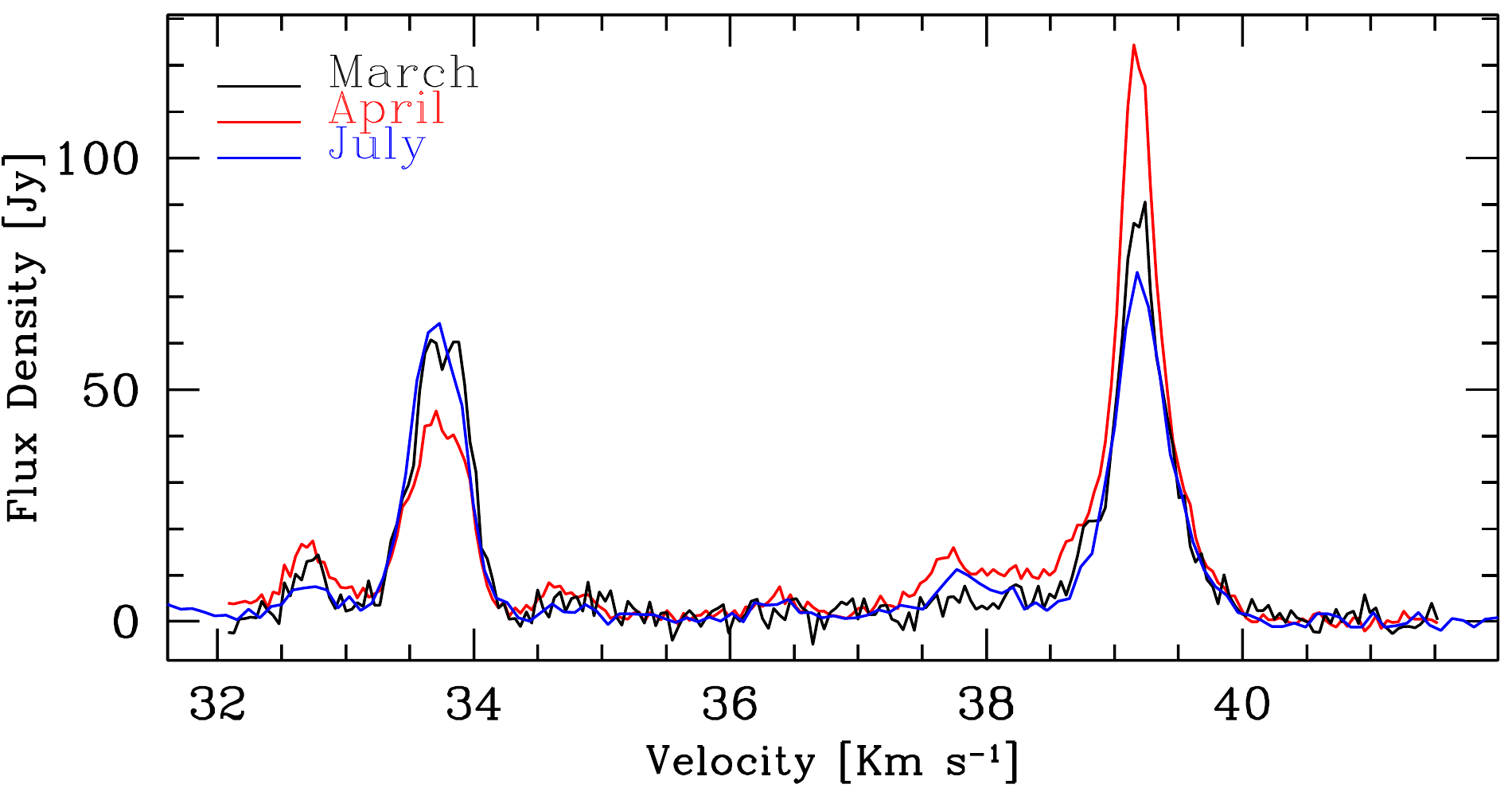}   
   \caption{6.7 GHz methanol maser spectra of the three epochs,
     obtained by summing all the pixels in the image for each channel.}
   \label{Fig:totspectra}
\end{figure*}

\begin{figure*}[htb]
   \centering
   \includegraphics[width=.3\textwidth]{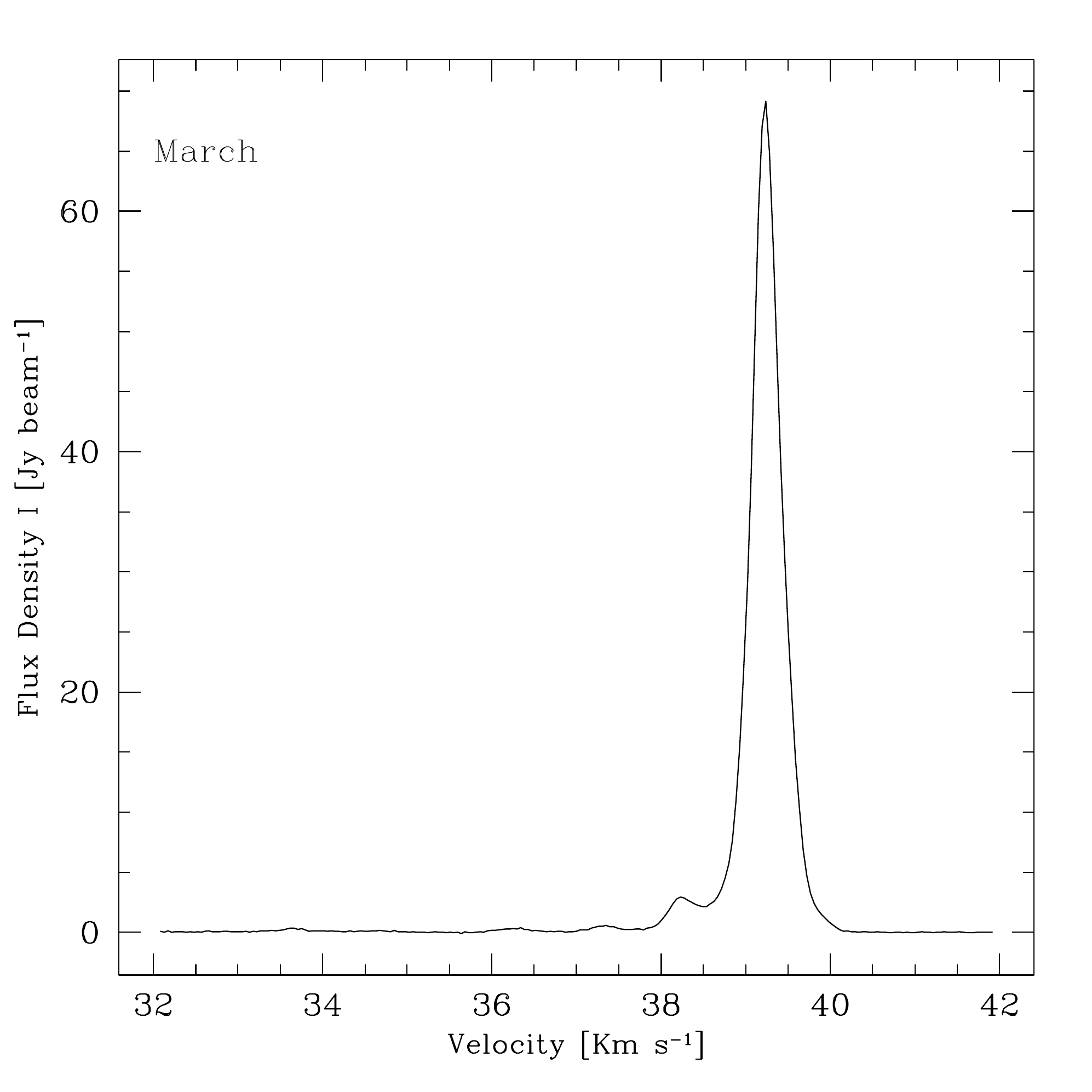}
   \includegraphics[width=.3\textwidth]{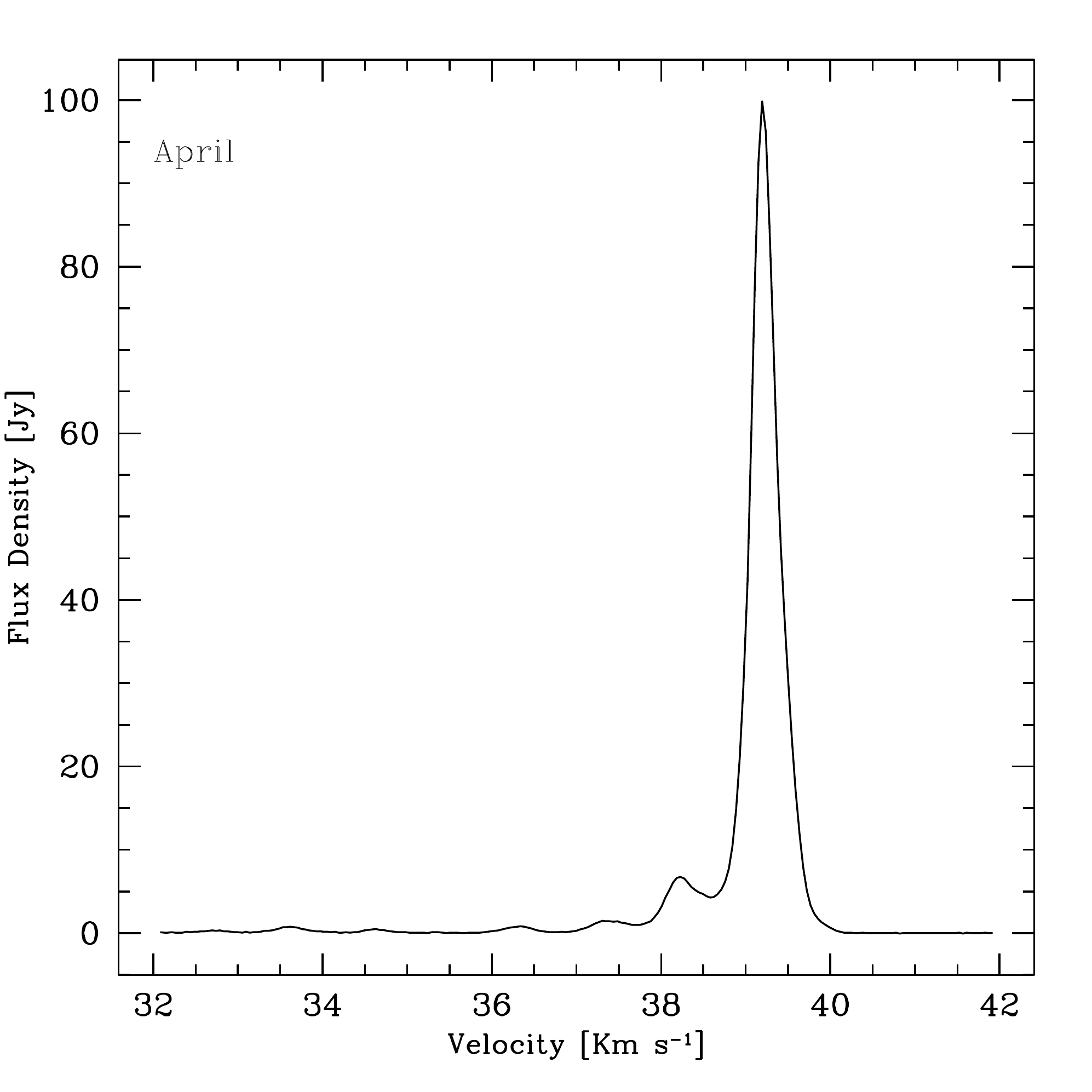}
   \includegraphics[width=.3\textwidth]{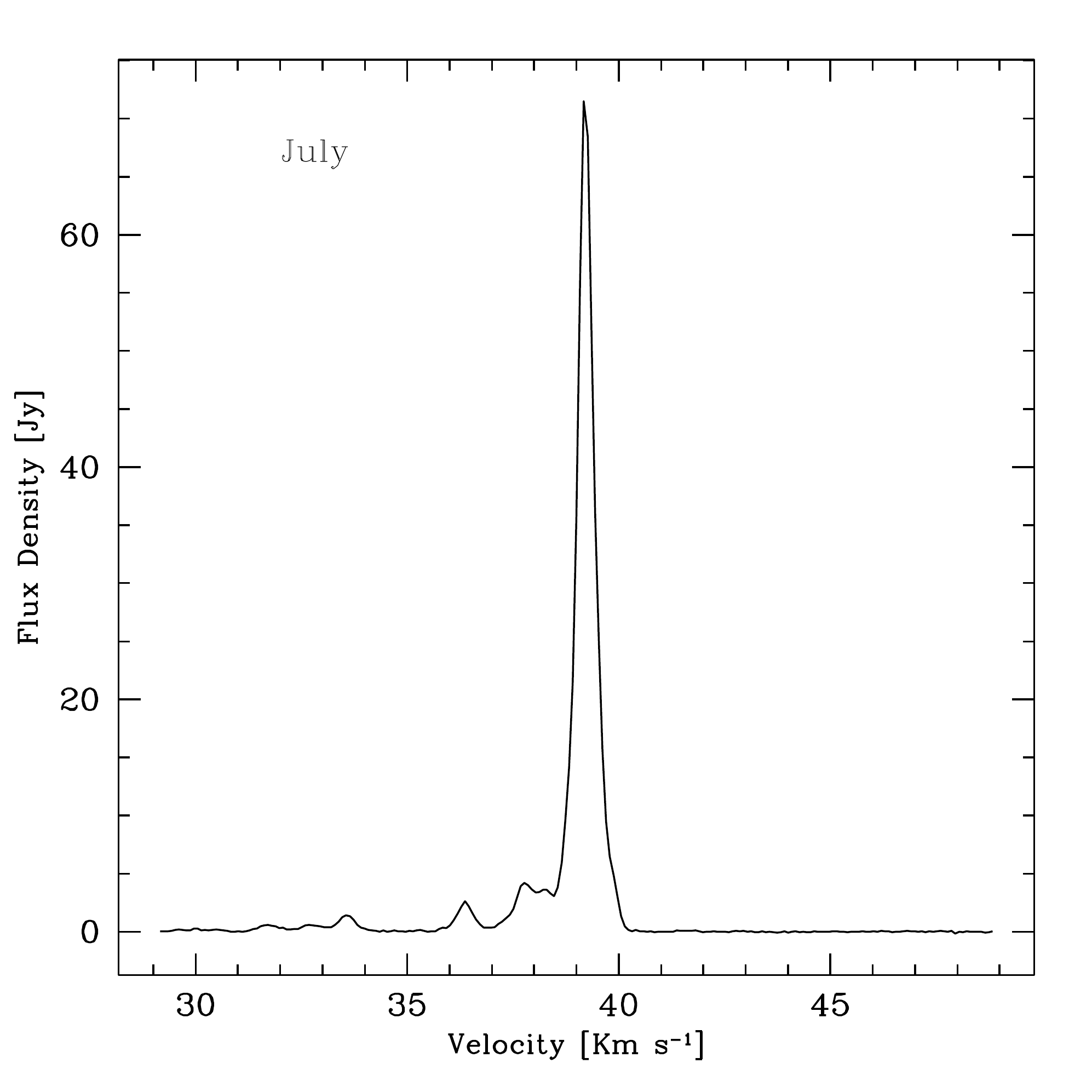}
   \caption{Spectra of the brightest maser feature F.01 in the three epochs.}
              \label{Fig:3spectra}
\end{figure*}

Self-calibration was performed on the brightest maser feature of
each epoch, using a solution interval of 30 seconds.
  The peak flux density of the brightest maser features was $\sim70$ Jy
beam$^{-1}$
in March, $\sim122$ Jy beam$^{-1}$ in
April
and $\sim75$ Jy beam$^{-1}$ in July, always at a velocity
$V_{lsr}$=39.2 km s$^{-1}$. The spectra of the three epochs, obtained
by summing all the pixels in the image for each channel, are plotted in Fig.~\ref{Fig:totspectra}, and
brightest maser feature spectra for the three epochs are shown in
Fig.~\ref{Fig:3spectra}.

We extracted the I, Q, U, and V cubes using
the task IMAGR (with an image size of $6.14\arcsec \times 6.14\arcsec$
and a cell size of $0.006 \arcsec$). As reported in
Table~\ref{tab:observational-details}, IRAS\,18089 was observed in full
polarisation mode only during March and April, while the July
observations were in dual circular polarisation only.
Thus, the linear polarisation calibration
was done only for these two epochs.  The RMS noise in the line-free
channels is reported in Table~\ref{tab:observational-details}; however the
noise increases in the channels with strongest maser
features where the noise is dominated by the dynamic range limitations.

In March, the noise
increased up to $\sim$0.2 Jy beam$^{-1}$ in I, $\sim$0.17 Jy
beam$^{-1}$ in Q, $\sim$0.13 Jy beam$^{-1}$in U, and $\sim$0.05 Jy
beam$^{-1}$ in V. In April, the noise reached $\sim$0.2 Jy beam$^{-1}$ in I, $\sim$0.25
Jy beam$^{-1}$ in Q, $\sim$0.15 Jy beam$^{-1}$ in U, and $\sim$0.07 Jy
beam$^{-1}$ in V. In July we only have maps in Stokes I and V, and in the channels with
the strongest features, the noise increased up to $\sim$0.12 Jy
beam$^{-1}$ and $\sim$0.06 Jy beam$^{-1}$, respectively.

In all epochs we estimated residual leakages of less than the RMS noise.

We combined the U and Q datacubes to produce cubes of polarised
intensity (POLI=$\sqrt{Q^2+U^2}$) and polarisation angle (POLA =
$1/2 \times \mathrm{atan} (U/Q)$). The error on POLA includes the
formal error due to the thermal noise
\citep{Wardle&Kronberg1974}. This error is given by
$\sigma_{POLA} = 0.5(\sigma_{P}/POLI)\times(180^{\circ}/\pi)$, where
$\sigma_{P}$ is the RMS of POLI. We compared the linear polarization
angles for 3C286 measured in our observations with the angles reported
by the NRAO in the Polarization Calibration
Database\footnote{www.vla.nrao.edu/astro/calib/polar/2008}. 3C286 is a
standard well-known calibrator with a stable polarisation angle of $33.0^{\circ}$
at 6.7 GHz. After the calibration, we found angles of $38^{\circ}\pm5^{\circ}$ and
$36^{\circ}\pm3^{\circ}$ in March and April respectively, which are consistent with
the angle given in the database.

\section{Results}
\label{sec:results}
The maser identification procedure follows the method already
described in \citet{Surcis2011w75n}. The \texttt{maser finder} code looks for
maser features inside the datacube, velocity channel by velocity
channel. The code recognises a maser spot when the signal to noise ratio (SNR) of the
candidate (using the local RMS) is greater than a predefined value. In
the case of IRAS\,18089, we adopted a lower limit of 8. A Gaussian fit
is performed for each maser spot using the AIPS task IMFIT; the code
generates a table containing parameters such as positions, velocities, and
peak flux densities for each spot. A maser feature is identified only
when the maser spots coincide spatially in at least three consecutive
velocity channels.
In March and April we found nine masers, while in July we found seven masers.

In Table~\ref{tab:maser_marchREF}, Table~\ref{tab:maser_aprilREF}, and
Table~\ref{tab:maser_julyREF} we present the 6.7 GHz methanol maser
features detected during the three epochs and that fulfil the criteria
described above. Not all features show up in all the three
epochs. Feature F.08 was observed only in March, while F.10 was
observed only in April, and F.11 only in July. The features maintain
the same names when they were observed across the three epochs.

We found a systematic shift of the positions of the brightest masers
between the three epochs but in each case this is less than the beam
size
. At the distance of IRAS\,18089,
proper motions in a few months would be negligible.
In the Tables, all maser positions are given as offsets relative to
the position of F.01 at that epoch. The absolute position of
  F.01 in March is
  $\alpha_{2000}=18^{\rm{h}}11^{\rm{m}}51.3954^{\rm{s}}$
  $\pm0.0001^{\rm{s}}$, $\delta_{2000}=-17^{\circ}31'29.92\arcsec$
  $\pm0.01\arcsec$ , in April it is
  $\alpha_{2000}=18^{\rm{h}}11^{\rm{m}}51.3929^{\rm{s}}$
  $\pm0.0003^{\rm{s}}$, $\delta_{2000}=-17^{\circ}31'29.86\arcsec$
  $\pm0.03\arcsec$ and in July it is
  $\alpha_{2000}=18^{\rm{h}}11^{\rm{m}}51.399^{\rm{s}}$
  $\pm0.002^{\rm{s}}$, $\delta_{2000}=-17^{\circ}31'29.93\arcsec$
  $\pm0.1\arcsec$. At $-17^\circ$ declination the errors are
dominated by the phase fluctuations. Other contributions to the
position accuracy include uncertainty in the position of the
telescope, noise-based errors in component fitting, and uncertainty in
the phase reference position.
The weighted
average position of the brightest maser F.01 is
$\alpha_{2000}=18^{\rm{h}}11^{\rm{m}}51.3949^{\rm{s}}$
$\pm0.0005^{\rm{s}}$, $\delta_{2000}=-17^{\circ}31'29.92\arcsec$
$\pm0.01\arcsec$.

In March, the offset of F.01 from the average position
was -33 mas in RA and 4.3 mas in Dec. In April the offset was -68.6 mas in RA and
64.0 mas in Dec. In July the offset was 16.0 mas in RA and -8.1 mas in Dec.  The peak
flux densities reported in the tables were obtained from a Gaussian
fit to the image of the brightest emission channel of the maser
feature. The error on the relative positions
reported in the tables was computed following \citet{Reid1988}.

No cross-hand polarisation products were correlated for the July
observation, and hence no linear polarisation values were obtained.
Therefore we could only perform the linear polarisation analysis  for
the March and April datasets.  Four features in March and six features
in April showed linear polarisation for which we measured the median
linear polarisation fraction ($P_{\rm{l}}$) and the median linear
polarisation angle ($\chi$) across the
spectrum.
The values for $P_{\rm{l}}$ and $\chi$ are
also reported in Table~\ref{tab:maser_marchREF} and
Table~\ref{tab:maser_aprilREF}. In our polarisation analysis we only
considered maser features with intensity $>$1 Jy.

All the masers that we  identified are plotted in Fig.~\ref{Fig:fieldtot}.  The bottom panels show a zoom of the regions
marked by the dashed grey boxes in the top panels, to better discern
between very close maser features. Each maser is represented by a
triangle. The different sizes of the triangles represent the
intensity, while the colours indicate the velocity of the maser
feature, following the scale reported in the colour bar. For
March and April only, the line segments mark the direction of the polarisation
angle for the maser features that show linear polarisation. Under the
assumption that the angle between the magnetic field and the line of
sight is $\theta > \theta_{crit}\sim 55^{\circ}$, where
$\theta_{crit}$ is the Van Vleck angle (e.g. \citealt{Surcis2011w75n}),
we considered the linear polarisation perpendicular to the magnetic
field. The vectors are also scaled logarithmically according to
$P_{\rm{l}}$ as reported in Tables~\ref{tab:maser_marchREF} and
\ref{tab:maser_aprilREF}. We also plot, in the bottom right
corners, the average direction of the resulting magnetic field
$\Phi_{B}$ obtained for two groups of masers as defined in
Sect.~\ref{sec:orientation}.
For the F.06 feature we tentatively detect a circular
polarisation signature and the magnetic field along the line of sight (see
Sect.~\ref{sec:strength}).


\begin {table*}[!p] 
  \caption []{Parameters of the 6.7-GHz methanol maser features detected in IRAS\,18089-1732 in March.} 
\begin{center}
\begin{tabular}{ l c c c c c c c c c }
\hline
\hline
\,\,\,\,\,(1) & (2)                 & (3)                  & (4)              & (5)            & (6)                           & (7)                     & (8)            & (9)          & (10)          \\
Maser         & RA\tablefootmark{a} & Dec\tablefootmark{a} & Peak flux        & $V_{\rm{lsr}}$ & $P_{\rm{l}}\tablefootmark{b}$ & $\chi\tablefootmark{b}$ & $\Delta_{V_L}$ & $P_{\rm{V}}$ & B$_{los}$     \\
              & offset              & offset               & Density(I)       &                &                               &                         &                &              &               \\ 
              & (mas)               & (mas)                & (Jy beam$^{-1}$) & (km $s^{-1}$)  & (\%)                          & ($\circ$)               & (km $s^{-1}$)  & ($\%$)       & (mG)          \\ 
\hline
F.01          & $ 0      \pm   0.13 $          & $ 0        \pm 0.78   $         & $69.99\pm 0.66$  & 39.24          & $8.9\pm1.4$                   & $-78\pm5$               & $-$            & $-$          & $-$           \\ 
F.02          & $ -34.90 \pm   0.21 $          & $ 14.11    \pm 1.26   $         & $8.53 \pm 0.13$  & 38.84          & $-$                           & $-$                     & $-$            & $-$          & $-$           \\ 
F.03          & $ 43.68  \pm   0.15 $          & $ -2.46    \pm 0.88   $         & $5.19 \pm 0.06$  & 37.75          & $-$                           & $ $                     & $-$            & $-$          & $-$           \\ 
F.04          & $ 28.92  \pm   0.18 $          & $ 15.60    \pm 1.11   $         & $3.23 \pm 0.04$  & 36.43          & $-$                           & $-$                     & $-$            & $-$          & $-$           \\ 
F.05          & $ 157.26 \pm   0.16 $          & $ 51.95    \pm 0.96   $         & $4.12 \pm 0.05$  & 34.67          & $-$                           & $-$                     & $-$            & $-$          & $-$           \\ 
F.06          & $ 1098.47\pm   1.07 $          & $ 1128.63  \pm 6.43   $         & $40.04\pm 0.40$  & 33.84          & $3.8\pm2.7$                   & $-50\pm28$              & $0.4$          & $0.8$        & $5.5 \pm 1.7$ \\ 
F.07          & $ 54.65  \pm   0.21 $          & $ 54.52    \pm 1.28   $         & $17.10\pm 0.27$  & 33.53          & $6.3\pm0.4$                   & $-31\pm1$               & $-$            & $-$          & $-$           \\ 
F.08          & $ 55.04  \pm   0.15 $          & $ 44.99    \pm 0.91   $         & $6.45 \pm 0.07$  & 32.74          & $9.4\pm0.3$                   & $-16\pm1$               & $-$            & $-$          & $-$           \\ 
F.09          & $ 937.80 \pm   0.16 $          & $ 1620.90  \pm 0.94   $         & $3.45 \pm 0.07$  & 32.70          & $-$                           & $-$                     & $-$            & $-$          & $-$           \\ 
\hline
\end{tabular} \end{center}
\tablefoot{
\tablefoottext{a}{The offsets are relative to the position of F.01 in
 March, i.e.\
 $\alpha_{2000}=18^{\rm{h}}11^{\rm{m}}51.3954^{\rm{s}} \pm 0.0001^{\rm{s}}$,
$\delta_{2000}=-17^{\circ}31'29.92\arcsec \pm 0.01\arcsec$.}\\
\tablefoottext{b}{$P_{\rm{l}}$ and $\chi$ are the median values of the
  linear polarisation fraction and the linear polarisation angle
  measured across the spectrum, respectively.}
 }
\label{tab:maser_marchREF}
\end{table*}


\begin{table*}[p!] 
\caption []{Parameters of the 6.7-GHz methanol maser features detected in IRAS\,18089-1732 in April.} 
\begin{center}
\begin{tabular}{ l c c c c c c c c c}
\hline
\hline
\,\,\,\,\,(1) & (2)                 & (3)                  & (4)              & (5)            & (6)                           & (7)                     & (8)              & (9)          & (10)          \\
Maser         & RA\tablefootmark{a} & Dec\tablefootmark{a} & Peak flux        & $V_{\rm{lsr}}$ & $P_{\rm{l}}\tablefootmark{b}$ & $\chi\tablefootmark{b}$ & $\Delta_{V_{L}}$ & $P_{\rm{V}}$ & $B_{los}$     \\
              & offset              & offset               & Density(I)       &                &                               &                         &                  &              &               \\ 
              & (mas)               & (mas)                & (Jy beam$^{-1}$) & (km s$^{-1}$)  & (\%)                          & ($\circ$)               & (km $s^{-1}$)    & ($\%$)       & (mG)          \\ 
\hline                                                                                                                                                  
F.01          &$ 0      \pm 0.10 $           &$ 0       \pm 0.59 $           & $122.29\pm 0.89$ & 39.20          & $8.5\pm1.3$                   & $-72\pm5$               & $-$              & $-$           & $-$           \\ 
F.02          &$ -36.16 \pm 0.14 $           &$ 24.21   \pm 0.85 $           & $13.38 \pm 0.14$ & 38.76          & $-$                           & $-$                     & $-$              & $-$           & $-$           \\ 
F.10          &$ 0      \pm 0.11 $           &$ -0.18   \pm 0.68 $           & $8.39  \pm 0.07$ & 38.23          & $6.9\pm0.7$                   & $-68\pm7$               & $-$              & $-$           & $-$           \\ 
F.03          &$ 48.23  \pm 0.09 $           &$ 18.08   \pm 0.55 $           & $17.57 \pm 0.12$ & 37.70          & $-$                           & $-$                     & $-$              & $-$           & $-$           \\ 
F.04          &$ 30.23  \pm 0.10 $           &$ 24.10   \pm 0.60 $           & $6.71  \pm 0.05$ & 36.39          & $9.4\pm1.2$                   & $-16\pm1$               & $-$              & $-$           & $-$           \\ 
F.05          &$ 162.97 \pm 0.08 $           &$ 72.11   \pm 0.51 $           & $11.22 \pm 0.07$ & 34.67          & $4.0\pm0.6$                   & $-46\pm1$               & $-$              & $-$           & $-$           \\ 
F.06          &$ 1110.26\pm 0.15 $           &$ 1127.96 \pm 0.87 $           & $42.78 \pm 0.46$ & 33.66          & $6.4\pm2.6$                   & $-16\pm12$              & $0.4$            & $0.3$        & $4.9 \pm 1.5$ \\ 
F.07          &$ 60.30  \pm 0.13 $           &$ 72.19   \pm 0.81 $           & $39.11 \pm 0.39$ & 33.58          & $8.8\pm0.1$                   & $-32\pm1$               & $-$              & $-$           & $-$           \\ 
F.09          &$ 941.32 \pm 0.41 $           &$ 1632.09 \pm 2.47 $           & $3.95  \pm 0.12$ & 32.70          & $-$                           & $-$                     & $-$              & $-$           & $-$           \\ 
\hline
\end{tabular} \end{center}
\tablefoot{
  \tablefoottext{a}{The offsets are relative to the position of F.01 in April, i.e.\ 
  $\alpha_{2000}=18^{\rm{h}}11^{\rm{m}}51.3929^{\rm{s}} \pm 0.0003^{\rm{s}}$
  $\delta_{2000}=-17^{\circ}31'29.86\arcsec \pm 0.03\arcsec$.} 
\tablefoottext{b}{$P_{\rm{l}}$ and $\chi$ are the median values of the linear polarisation fraction
  and the linear polarisation angle measured across the spectrum, respectively.}
}
\label{tab:maser_aprilREF}          
\end{table*}


\begin {table*}[p!] 
\caption []{Parameters of the 6.7-GHz methanol maser features detected in IRAS\,18089-1732 in July.} 
\begin{center}
\begin{tabular}{ l c c c c }
\hline
\hline
\,\,\,\,\,(1) & (2)                 & (3)                  & (4)              & (5)            \\ 
Maser         & RA\tablefootmark{a} & Dec\tablefootmark{a} & Peak flux        & $V_{\rm{lsr}}$ \\
              & offset              & offset               & Density(I)       &                \\
              & (mas)               & (mas)                & (Jy beam$^{-1}$) & (km $s^{-1}$)  \\
\hline
F.01          & $ 0       \pm   0.01   $        & $ 0       \pm 0.07  $            & $74.89\pm 0.06$  & 39.18          \\    
F.03          & $ 45.05   \pm   0.23   $        & $ -1.50   \pm 1.37  $            & $10.06\pm 0.17$  & 37.77          \\   
F.05          & $ 156.27  \pm   0.28   $        & $ 53.67   \pm 1.65  $            & $3.77 \pm 0.08$  & 34.70          \\     
F.06          & $ 1104.28 \pm   0.03   $        & $ 1123.61 \pm 0.17  $            & $47.90\pm 0.10$  & 33.73          \\     
F.07          & $ 56.79   \pm   0.08   $        & $ 60.12   \pm 0.49  $            & $18.42\pm 0.11$  & 33.56          \\     
F.09          & $ 938.19  \pm   0.50   $        & $ 1620.49 \pm 3.00  $            & $3.25 \pm 0.12$  & 32.68          \\     
F.11          & $ 59.70   \pm   0.32   $        & $ 35.35   \pm 1.94  $            & $2.47 \pm 0.06$  & 30.05          \\    
\hline
\end{tabular} \end{center}
\tablefoot{
  \tablefoottext{a}{The offsets are relative to the position of F.01 in July, i.e.\
    $\alpha_{2000}=18^{\rm{h}}11^{\rm{m}}51.399^{\rm{s}} \pm 0.002^{\rm{s}}$,
    $\delta_{2000}=-17^{\circ}31'29.93\arcsec \pm 0.1\arcsec$.}
}
\label{tab:maser_julyREF}
\end{table*}

\section{Discussion}
\label{sec:discussion}

\subsection{Maser distribution and kinematics}
\label{sec:distribution}
\citet{Goedhart2009} presented a spectrum of the 6.7 GHz methanol
maser emission of IRAS\,18089. The spectrum is divided in two distinct
blocks around two main peaks: one located at a velocity of 33.7
km~s$^{-1}$ and another at 39.2 km s$^{-1}$.  In our observations, we
found a similar spectrum, with all our maser features grouped around
those two values (see
Tables~\ref{tab:maser_marchREF}--\ref{tab:maser_julyREF}): as shown in
the Tables, the velocity shifts are within the channel width, apart
from F.06 which presents a shift of $\sim$0.2 km~s$^{-1}$, probably
due to a blend of two components (see Sect.~\ref{sec:strength}). Since
the same velocities occur in all three of our epochs, we divided the
masers in two groups: a blue group spanning a velocity range from 30.0
to 36.4 km~s$^{-1}$ (containing F.04, F.05, F.06, F.07, F.08, F.09,
and F.11), and a red group from 37.7 to 39.2 km~s$^{-1}$ (containing
F.01, F.02, F.03, and F.10). The blue group presents a velocity that
is similar to the v$_{lsr}$ of the compact core ($\sim$33.8
km~s$^{-1}$, \citealt{Beuther2005,Leurini2007}) while the red group is
closer to the velocity of the CO outflow ($\sim$40 km~s$^{-1}$,
\citealt{Beuther2010}).

\begin{sidewaysfigure*}[p]
  \includegraphics[width=0.33\columnwidth, bb=0 16 556 472,clip]{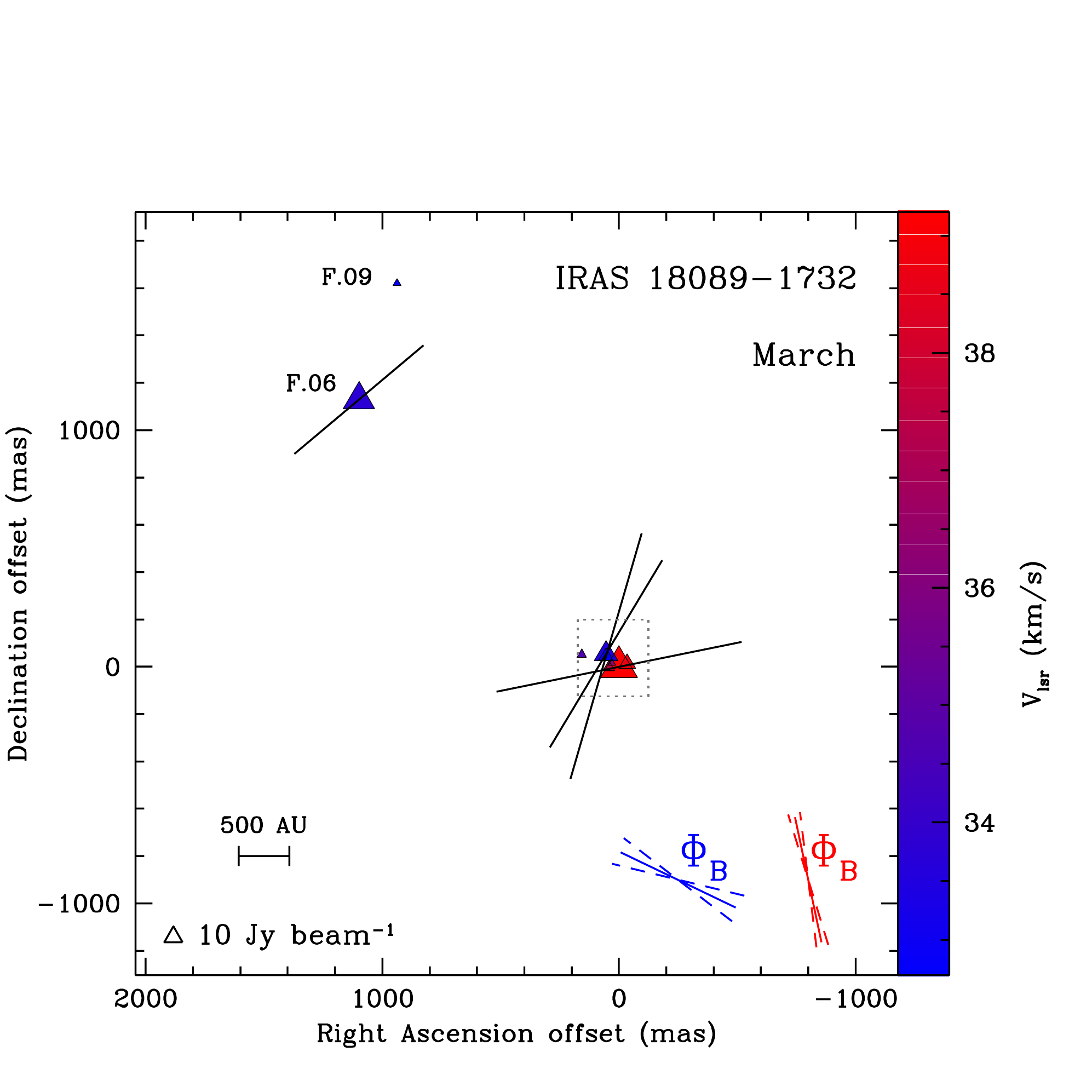}
  \includegraphics[width=0.33\columnwidth, bb=0 16 556 472,clip]{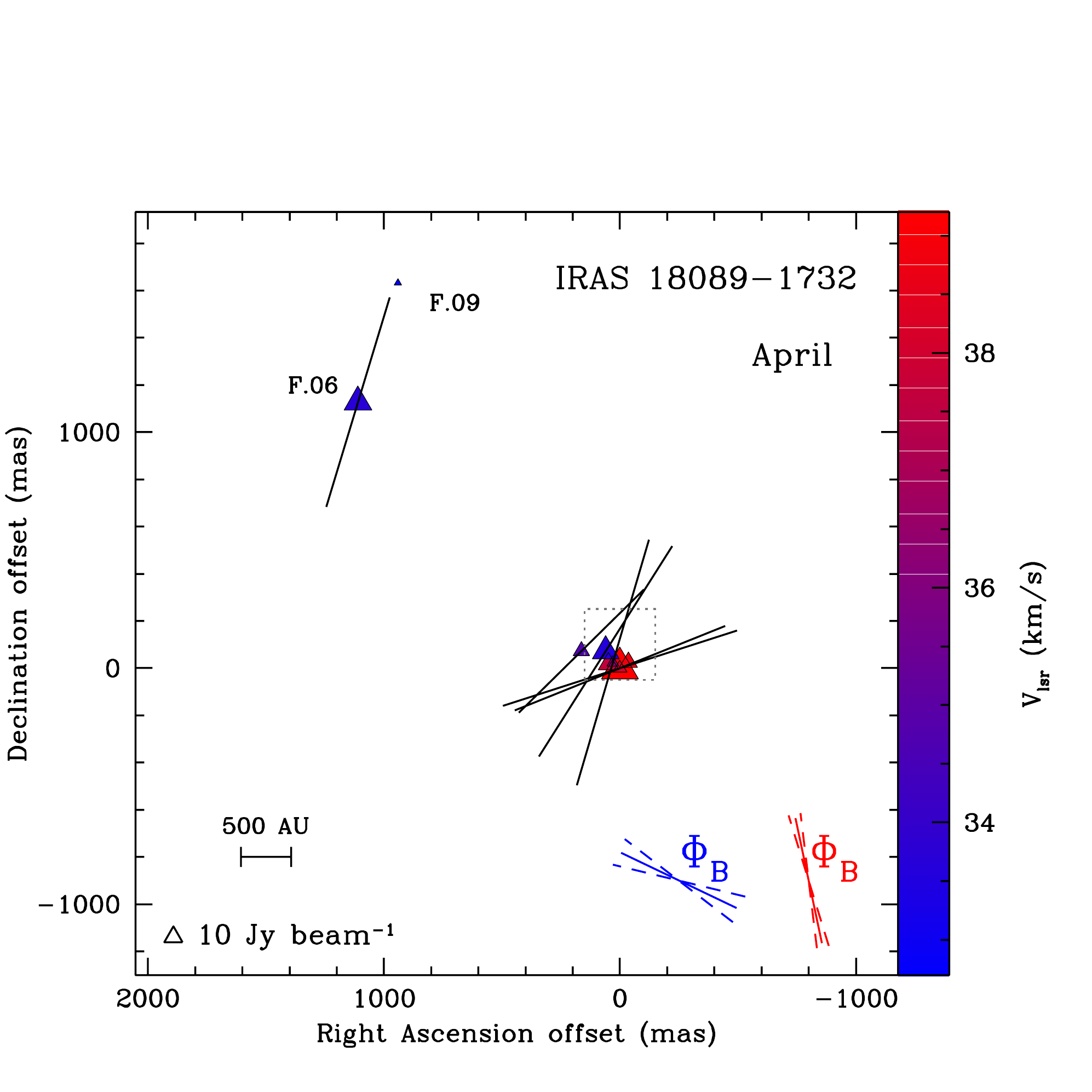}
  \includegraphics[width=0.33\columnwidth, bb=0 16 556 472,clip]{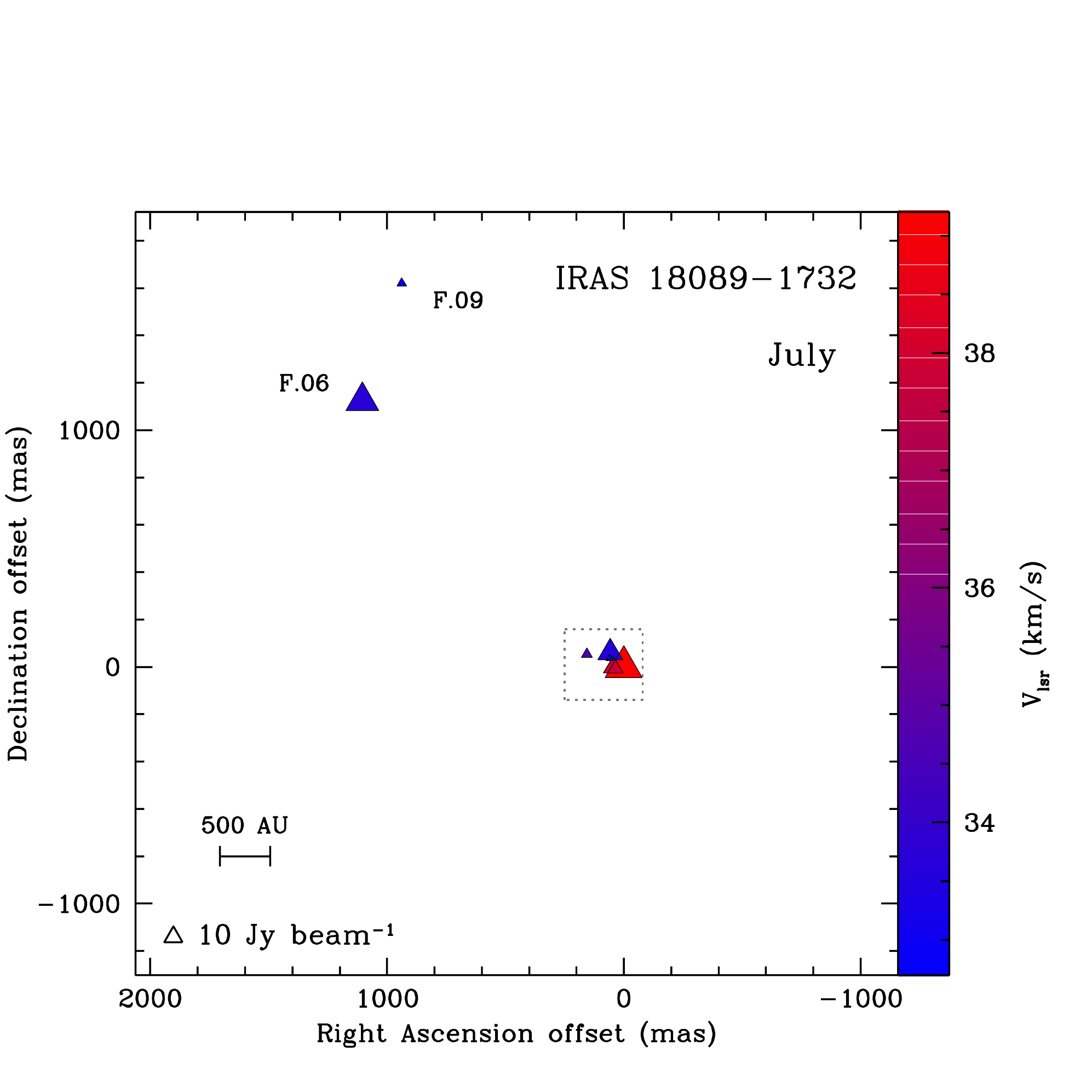}\\ 
  \includegraphics[width=0.33\columnwidth, bb=0 16 556 472,clip]{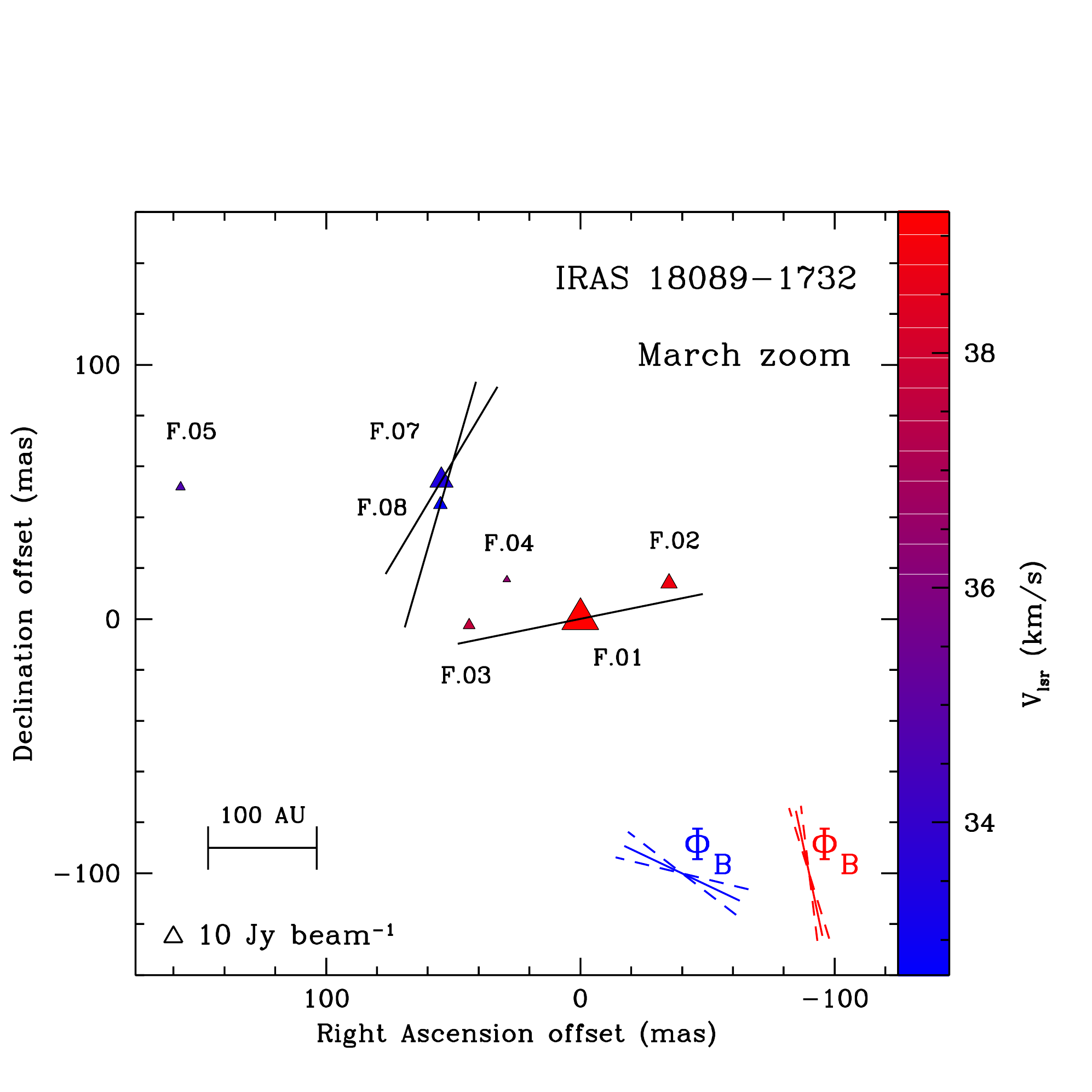}
  \includegraphics[width=0.33\columnwidth, bb=0 16 556 472,clip]{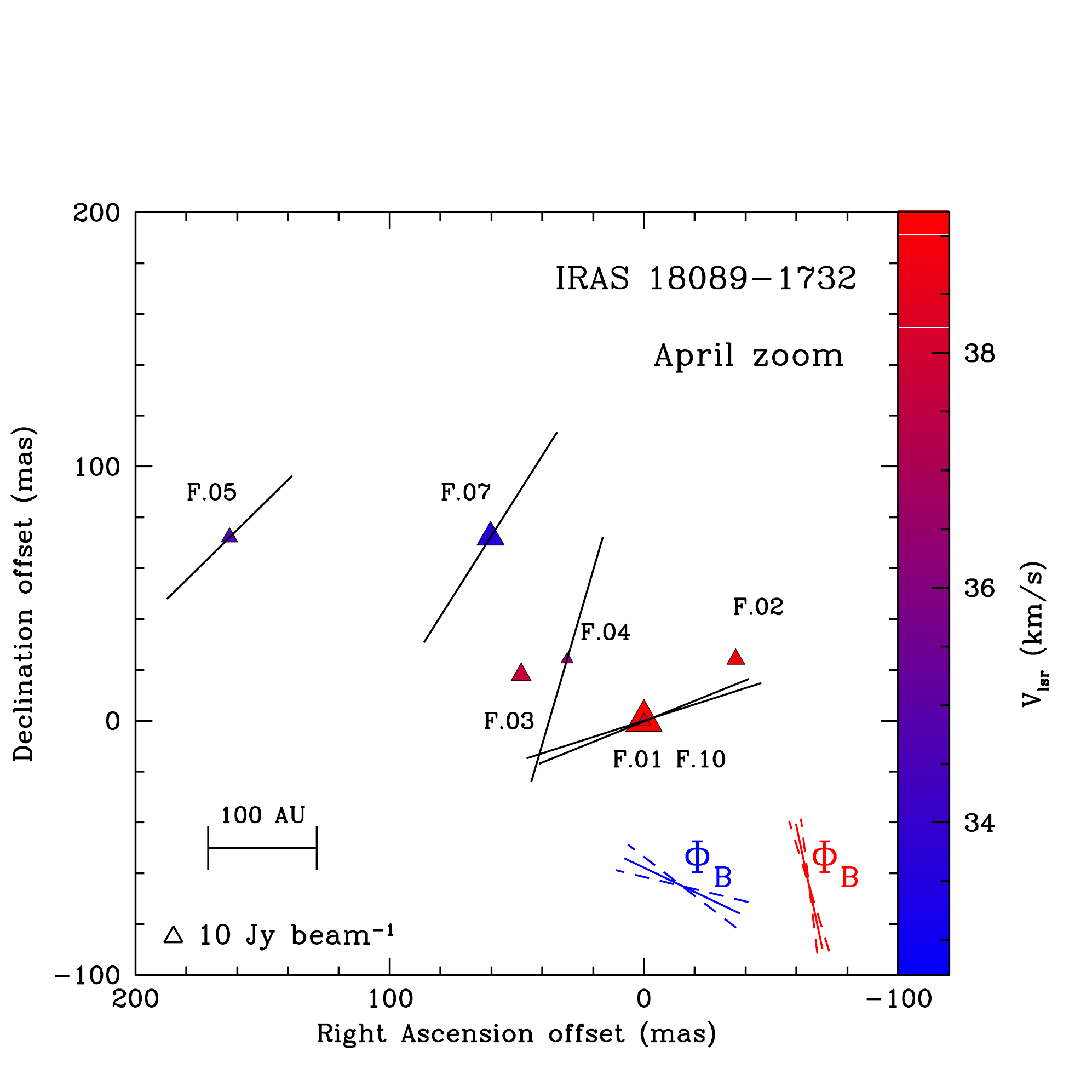}
  \includegraphics[width=0.33\columnwidth, bb=0 16 556 472,clip]{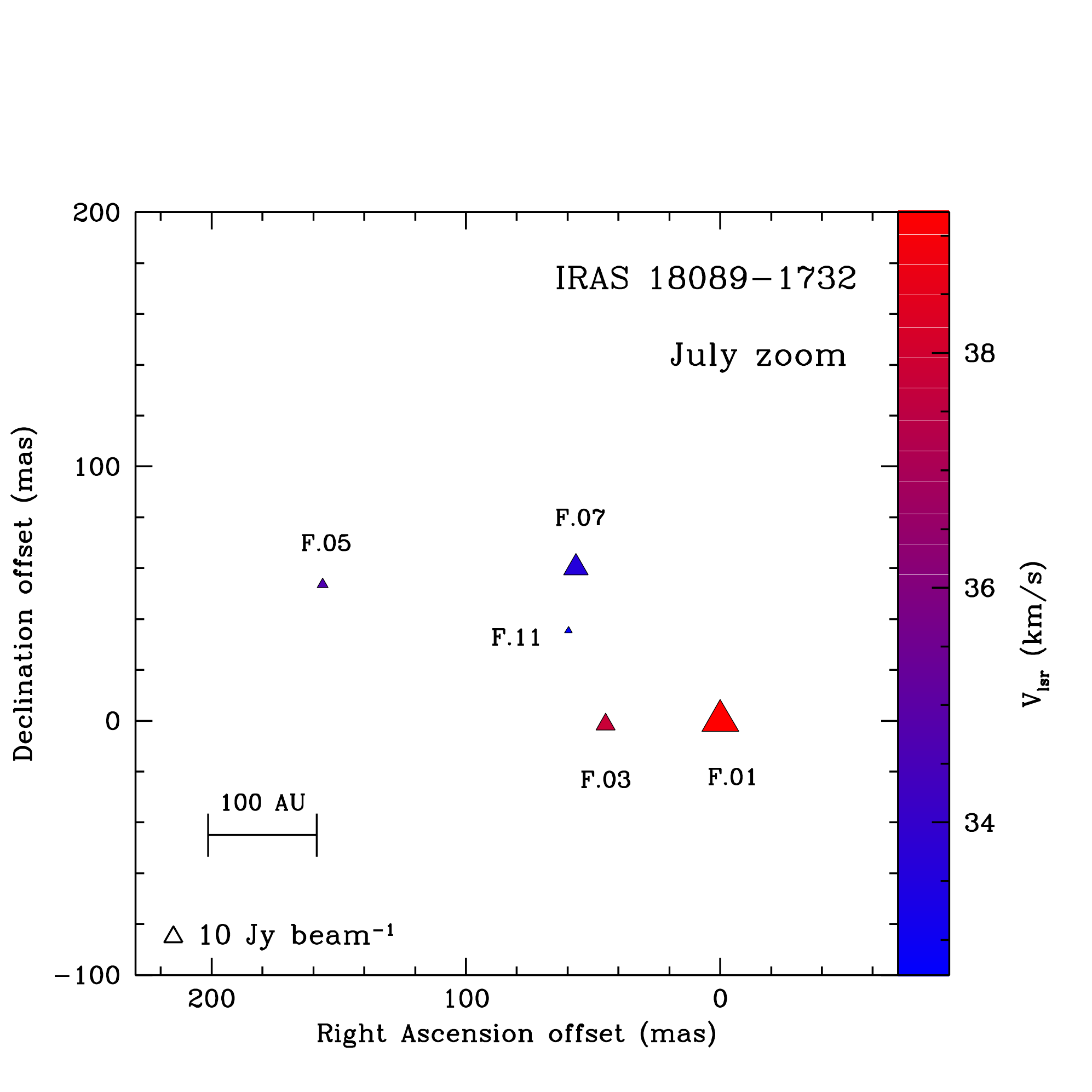}\\
   \caption{Masers identified in March (left), April (centre) and July (right) as listed in
     Table~\ref{tab:maser_marchREF}--\ref{tab:maser_julyREF}. The bottom panel shows a zoom of
     the region marked by the dashed grey boxes in the top
     panels. Each maser is represented by a triangle. The different
     sizes of the triangles represent the intensity, while the colours
     indicate the velocity of the maser feature, according to the scale
     reported in the colour bar. Line segments mark the direction of
     the polarisation angle for the maser features that show linear
     polarisation. The average direction of the resulting magnetic
     field $\Phi_{B}$ obtained for two groups of masers as defined in
     Sect.~\ref{sec:orientation} is indicated in the bottom right
     corners of each panel.The July observations were in dual circular polarisation only.}
   \label{Fig:fieldtot} 
\end{sidewaysfigure*}

\subsection{Orientation of the magnetic field }
\label{sec:orientation}

\begin{figure*}[htbp]
   \centering
   \includegraphics[width=.9\columnwidth]{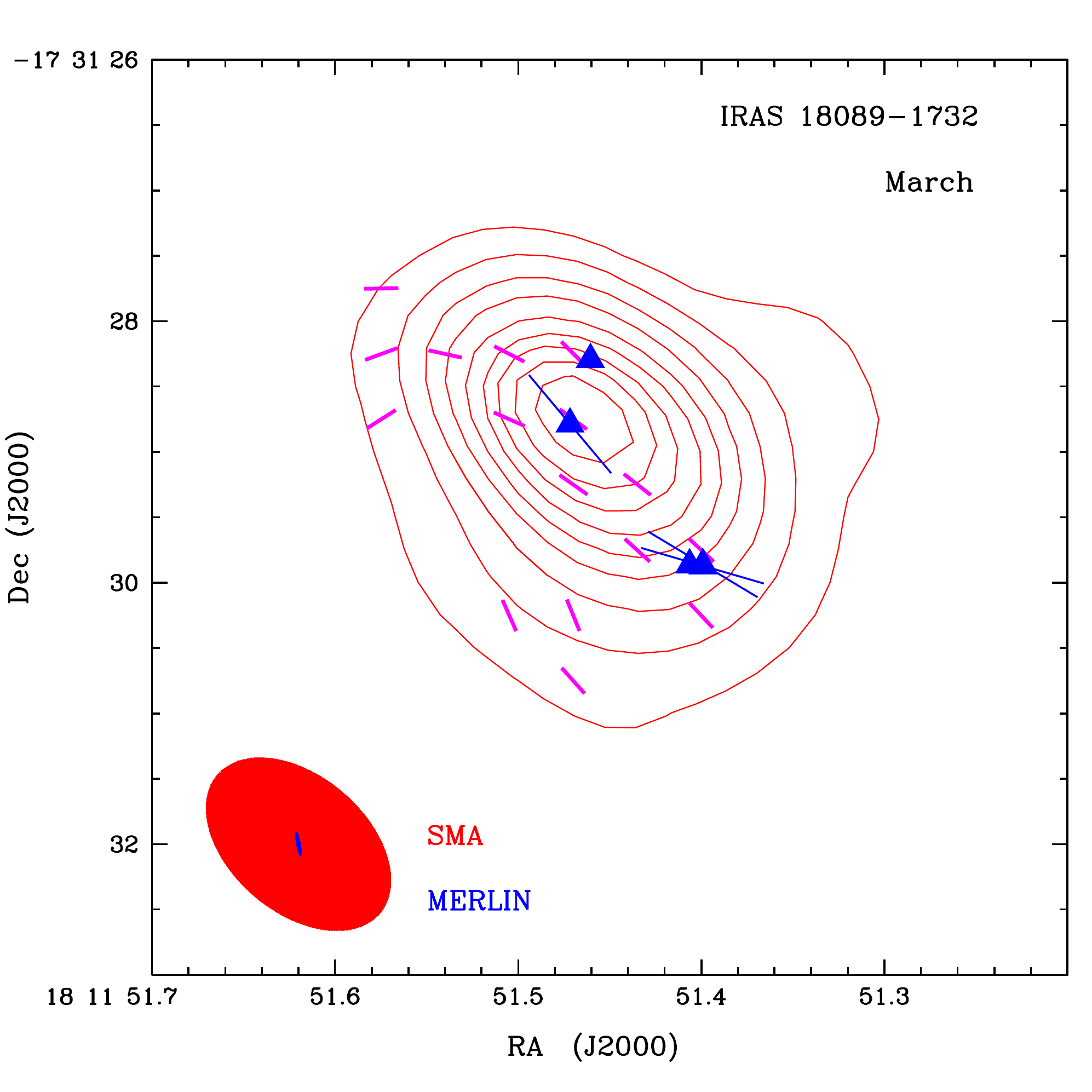}
   \includegraphics[width=.9\columnwidth]{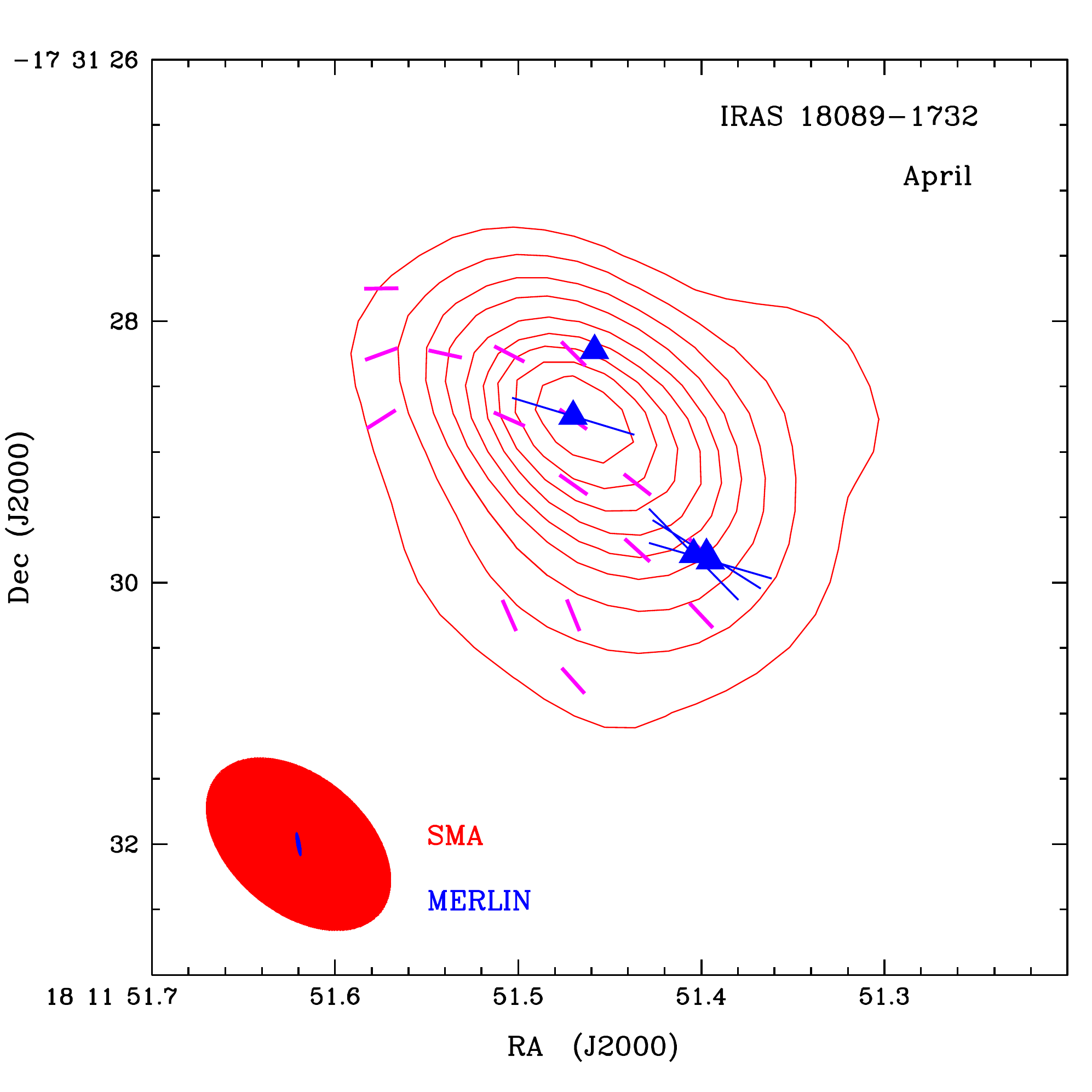}
   \caption{Masers in the blue group (blue triangles and blue
     segments) superimposed on the integrated I image of the dust
     continuum emission observed by \citet{Beuther2010} at 880~$\mu$m
     with SMA (red contours; the contours are drawn in $10 \sigma$
     steps). The magenta line segments show the magnetic field
     orientation obtained by linearly polarised dust emission
     \citep{Beuther2010}. The blue segments represent the magnetic
     field orientation obtained by our linearly polarised methanol
     maser emission (Sect.~\ref{sec:orientation}); therefore the
     magnetic field follows the same direction indicated by the dust
     emission.  The red and blue ellipses show the beams of SMA
     ($1.65\arcsec \times 1.05\arcsec$, position angle $51^{\circ}$) and
     MERLIN, respectively. Left panel: March; right panel: April.}
              \label{Fig:overlappingBeuther2010}
\end{figure*}

The difference in velocity and the separation of the two groups is also
preserved in the orientation of the polarisation vector of each maser
in both March and April. The two groups of masers show ordered linear
polarisation vectors.  The blue (B) group has a weighted average angle of
$\langle\chi_{B,M}\rangle = -24^{\circ} \pm 8^{\circ}$ in March (M) and
$\langle\chi_{B,A}\rangle= -31^{\circ} \pm 12^{\circ}$ in April (A).
The red (R) group has only one maser feature presenting linear polarisation in March with an
angle $\chi_{R,M}= -78^{\circ} \pm 5^{\circ}$, while in April the
weighted polarisation angle is
$\langle\chi_{R,A}\rangle = -70^{\circ} \pm 2^{\circ} $. Therefore
the two groups show distinct polarisation angles as well as velocities.

The presence of these two groups of masers, showing two characteristic
orientations of the polarisation angle and two velocity ranges, can be
interpreted as a distinct signature of masers originating in two
different regions around the protostar.  Under this hypothesis, the
two groups of masers probe the magnetic field morphology in two
different regions. In Fig.~\ref{Fig:fieldtot} we plot the two directions of the
resulting magnetic field $ \Phi_{B} $, assuming a magnetic field
perpendicular to the polarisation angles (see
Sect.~\ref{sec:results}).  The masers in the blue group have an
orientation on the plane of the sky
$\Phi_{B}^{disc}= +62^\circ\pm3^\circ$, probing the magnetic field
close to the disc. Conversely, the masers in the red group have an
orientation on the plane of the sky
$\Phi_{B}^{outflow}= +14^\circ\pm4^\circ$, probing the magnetic field
close to the base of the outflow.

\citet{Beuther2010} detected dust polarised emission describing the
IRAS\,18089 disc, and the accompanying line features have a velocity
similar to that identified for the blue group of masers.  In
Fig.~\ref{Fig:overlappingBeuther2010} we overplot the blue group
masers on the Stokes I continuum emission observed by
\citet{Beuther2010}.  The red and blue ellipses represent the SMA and
MERLIN beams, respectively. In the left panel we plot the masers
identified in March, while in the right panel those identified in
April. The blue lines show the direction of the magnetic field,
obtained by our maser linear polarisation analysis. The magenta
segments identify the magnetic field direction as found by
\citet{Beuther2010} from the polarised dust emission. This suggests
that the small-scale magnetic field probed by the masers is consistent
with the large-scale magnetic field traced by the dust. Therefore we
can conclude that the magnetic field structures remain consistent over
 many orders of magnitude in scales.

 \citet{Beuther2010} also found CO(3-2) emission tracing the outflow
 at a velocity consistent with that found for our red group
 masers. However, the CO(3-2) outflow is located too far away
 ($\sim4\arcsec$ from the peak of CO(3-2) and $\sim1.5\arcsec$ from
 the lowest contour) from the red group masers, so it is not tracing
 the same gas.  \citet{Beuther2004b} observed several other molecules
 tracing the outflow including SiO(5-4) and H$_2$S. Both SiO(5-4) and
 H$_2$S present a more extended emission than that of CO(3-2), that
 overlaps with the area where our red group masers lie. The channel
 maps present by \citet{Beuther2004, Beuther2004b} show that both
 SiO(5-4) and H$_2$S have a velocity consistent with our red group
 masers.  Our hypothesis is that the red group masers are tracing the
 magnetic field responsible for, or related to, the outflow.

\subsection{Maser variability }
\label{sec:variability}
\begin{figure*}
   \centering
   \includegraphics[width=.9\columnwidth]{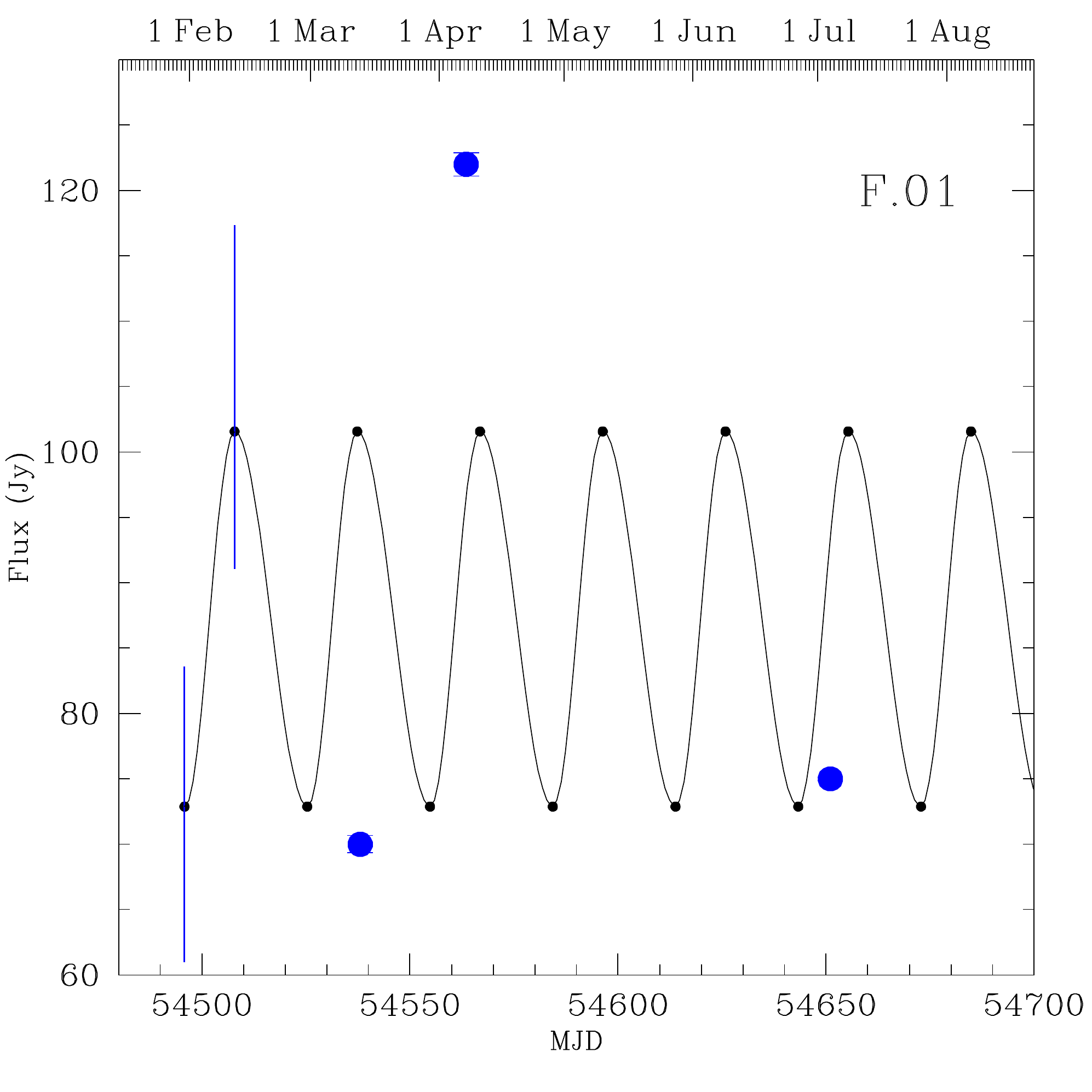}
   \includegraphics[width=.9\columnwidth]{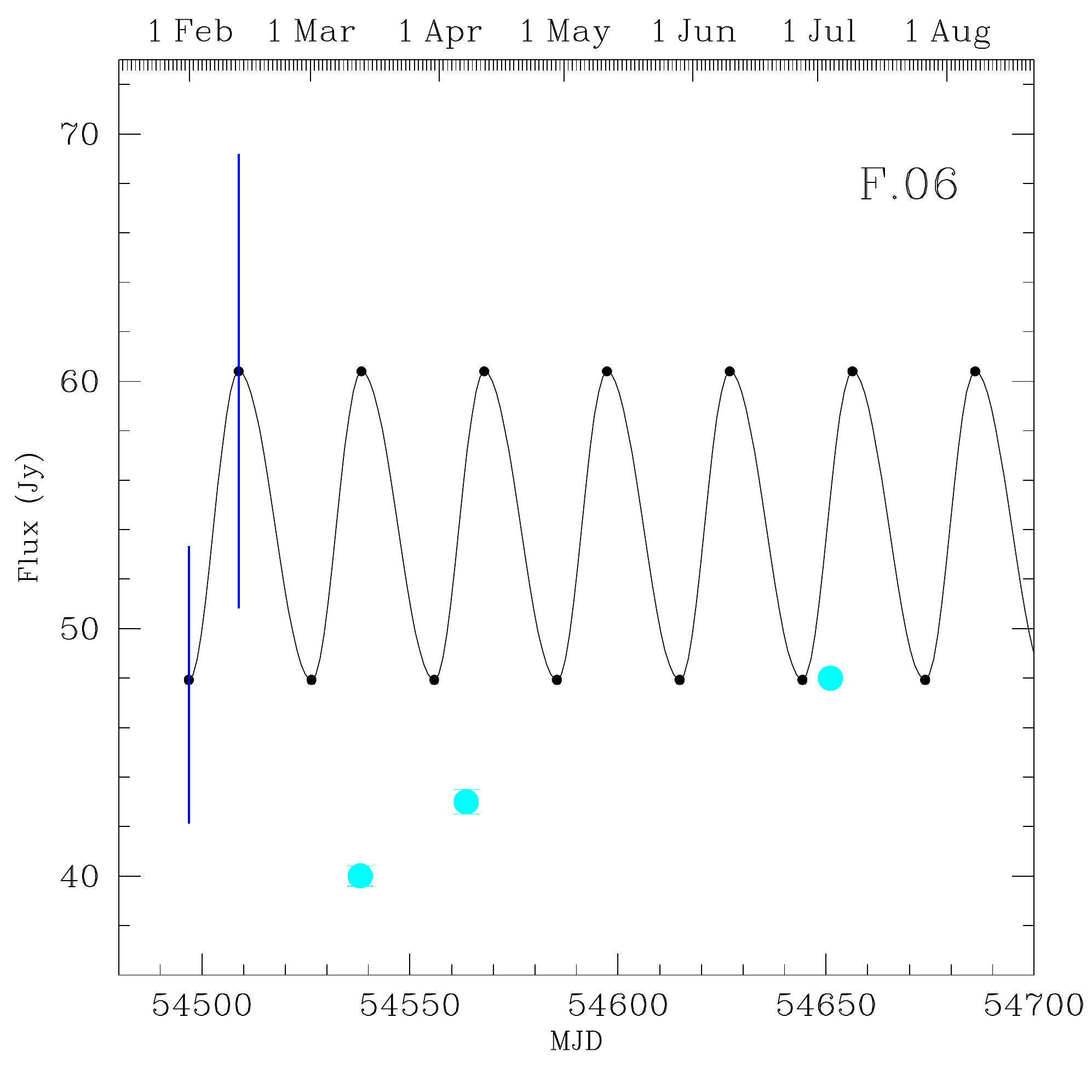}
   \caption{Flux density versus time for the features F.01 (left) and
     F.06 (right). The black curve illustrates the expected
     variability of the maser feature according to
     \citet{Goedhart2009}; the predicted times of maxima and minima
     are reported in Table~\ref{tab:minmax}. The blue bars indicate
     the range of the intensities between the highest maximum and the lowest maximum and
     between the highest minimum and lowest minimum.}
              \label{Fig:variability}
\end{figure*}

\citet{Goedhart2009} observed a quasi-periodic variability in
IRAS\,18089. During 9 years of observations they registered a light
curve presenting regular minima but with variable amplitude of flares
and phase of the peak. The most likely period is 29.5$\pm$0.1
days. The feature at 39.2 km~s$^{-1}$ takes roughly 12 d on average to
reach the maximum after a minimum. Extrapolating from the last minimum
observed by \citet{Goedhart2009} in 2007, we can predict the dates of
the maxima and minima for the features in common with our observations
(see Table~\ref{tab:minmax} and Fig.~\ref{Fig:variability}). The uncertainties reported in
Table~\ref{tab:minmax} have been calculated by multiplying the
uncertainty on the period (0.1 d) by the square root of the number of
elapsed periods between the last minimum observed by
\citet{Goedhart2009} and our observations (12--16 periods).  Our March
observations took place one day after a predicted maximum of F.01,
whilst the April and July observations took place 2--3 and 3--4 days
beforehand, respectively.

\begin{table*}[htbp]
  \caption[]{Predicted days and hours of minima and maxima in the
    light curve of the 39.2 km~s$^{-1}$ feature F.01, extrapolated
    from the light curve in \citet{Goedhart2009}. The predicted minima
    and maxima for F.02 can be obtained adding one day to the dates in
    the table.}
  \label{tab:minmax}
  \centering
  \begin{tabular}{ccc}
    \hline\hline
    Expected        & Expected & Uncertainty            \\
    Minima          & Maxima   & (days)                  \\
    \hline
29-02-2008 \ 07:30 & 12-03-2008 \ 07:30  &  $\pm$ 0.3  \\
29-03-2008 \ 19:30 & 10-04-2008 \ 19:30  &  $\pm$ 0.3  \\
28-04-2008 \ 07:30 & 10-05-2008 \ 07:30  &  $\pm$ 0.4  \\
27-05-2008 \ 19:30 & 08-06-2008 \ 19:30  &  $\pm$ 0.4  \\
26-06-2008 \ 07:30 & 08-07-2008 \ 07:30  &  $\pm$ 0.4  \\
25-07-2008 \ 19:30 & 06-08-2008 \ 19:30  &  $\pm$ 0.4  \\
    \hline           
  \end{tabular}          
\end{table*}

\begin{table*}[htbp]
  \caption[]{Timetable of our MERLIN observations.}
  \label{tab:ourobs}
  \centering
  \begin{tabular}{cc}
    \hline\hline
    Starting time          & Ending time   \\
    \hline
13-03-2008 \ 03:16 & 13-03-2008 \ 10:27   \\
07-04-2008 \ 01:49 & 07-04-2008 \ 09:07  \\
08-04-2008 \ 01:49 & 08-04-2008 \ 09:07   \\
04-07-2008 \ 19:32 & 05-07-2008 \ 02:29   \\
    \hline           
  \end{tabular}          
\end{table*}

In Fig.~\ref{Fig:totspectra}, the peaks around 33.8 and 39.2
km~s$^{-1}$ are dominated by F.06 and F.01, respectively, which from
this plot and Tables~\ref{tab:maser_marchREF}--\ref{tab:maser_julyREF}
can be seen to vary strongly in peak intensity. As indicated in
Fig.~\ref{Fig:variability}, left panel, F.01 is brightest in April, at
$\sim 122$ Jy~beam$^{-1}$, close to a predicted peak, but the March
and July observations, although also close to predicted peaks, have
flux densities of 70--75 Jy~beam$^{-1}$, closer to the values at
minimum reported by \citet{Goedhart2009}.  They found that the period
of F.06 lags F.01 by 1 day, so we would also expect to see F.06 close
to maximum. From Fig.~\ref{Fig:variability}, right panel, we instead
measured flux densities of 40--47 Jy~beam$^{-1}$ which are closer to
the minimum values found by \citet{Goedhart2009}. This is not entirely
unexpected, since \citet{Goedhart2009} pointed out that the times of
maxima are not very regular.  Another possibility is that these
features have dropped in flux density (since we did not observe at the
predicted minima, it is possible that the features had much lower
fluxes then).

Following \citet{Goedhart2005} and \citet{Goedhart2009}, a possible
way to understand the delay between the peaks is by analysing the
projected distances of the masers and the light travel times. The
projected distance between F.01 and F.06 is $\sim$3700 au and the
light travel time across this length is
$\sim$20~d. \citet{Goedhart2009} observed delays also for other
features. Between F.01 and F.02 and F.01 and F.03 the expected delays
are -1.7 and -3.3 days, respectively. The projected distance between
F.01 and F.02 features is $\sim$90~au and the corresponding light
travel time is $\sim$0.5~d. In the case of F.01 and F.03 the projected
distance is 100~au, equivalent to $\sim$0.6 light days. Between F.01
and F.05 the observed delay is 1.7~d, and the projected distance is
$\sim$380~au corresponding to $\sim$2.3 light days.  The features with
a negative delay coincide with our red group, while the features with
positive delays coincide with the blue group. If we assume that the
variability of all the features is due to the same pumping source, it
should be located on a plane between F.01 and F.06. F.02 and F.03
should be located closer to the pumping source than F.01. However,
because of the degeneracy of the positions along the line of sight,
many configurations are possible and at the current status we cannot
determine a single three-dimensional model of the region.

\citet{Walsh1998} provided a map of IRAS\,18089 masers and obtained
relative positions with an accuracy of $\sim0.05\arcsec$ and absolute
positions accurate to $\sim1\arcsec$. Through the study of
position-velocity (P-V) diagrams, no evidence of peculiar structure
within the maser site was found. We compared the relative positions of
the masers observed by \citet{Walsh1998} assuming the same absolute
position for the brightest features at 39.2 km~s$^{-1}$. For an easy
comparison, we report, in Table~\ref{tab:conversion}, the maser features
observed in our MERLIN observations, in \citet{Walsh1998}, and in
\citet{Goedhart2009}.
In our three epochs we observed almost all the features already
detected by \citet{Walsh1998}, at positions within their reported
uncertainty of 50 mas. However, there are some exceptions: our F.04,
F.07, F.08, F.09 and F.10 weren't detected by \citet{Walsh1998}, while
we didn't detect features D, G, and H at 32.7 and 31.6 km~s$^{-1}$.

\begin{table*}
  \caption[]{Conversion table of the maser features, between our MERLIN observations, \citet{Walsh1998}, and  \citet{Goedhart2009}. Velocities in km~s$^{-1}$ are also reported for comparison.}
  \label{tab:conversion}
  \centering
  \begin{tabular}{ccc}
    \hline\hline
    This paper   & \citet{Walsh1998} & \citet{Goedhart2009} \\
    \hline
    F.01 \   39.2 & A \  39.2          & 39.2                 \\
    F.02 \   38.8 & B \  38.7          & 38.7                 \\
    F.03 \   37.7 & C \  37.6          & 37.7                 \\
    F.04 \   36.4 & ---                & \\
     ---          & D \  36.6          & 36.6                 \\
    F.05 \   34.7 & E \  34.6          & 34.6                 \\
    F.06 \   33.7 & F \  33.6          & 33.6                 \\
     ---          & G \  32.7          & 32.7                 \\
     ---          & H \  31.6          & 31.6                 \\
    F.07 \   33.6 & ---                & ---                  \\
    F.08 \   32.7 & ---                & ---                  \\
    F.09 \   32.7 & ---                & ---                  \\
    F.10 \   38.2 & ---                & ---                  \\
    F.11 \   30.0 & J \  30.0          & 30.0                 \\  
    \hline           
  \end{tabular}          
\end{table*}

\subsection{Strength of the magnetic field }
\label{sec:strength}

\begin{figure*}
   \centering
   \includegraphics[width=.9\columnwidth]{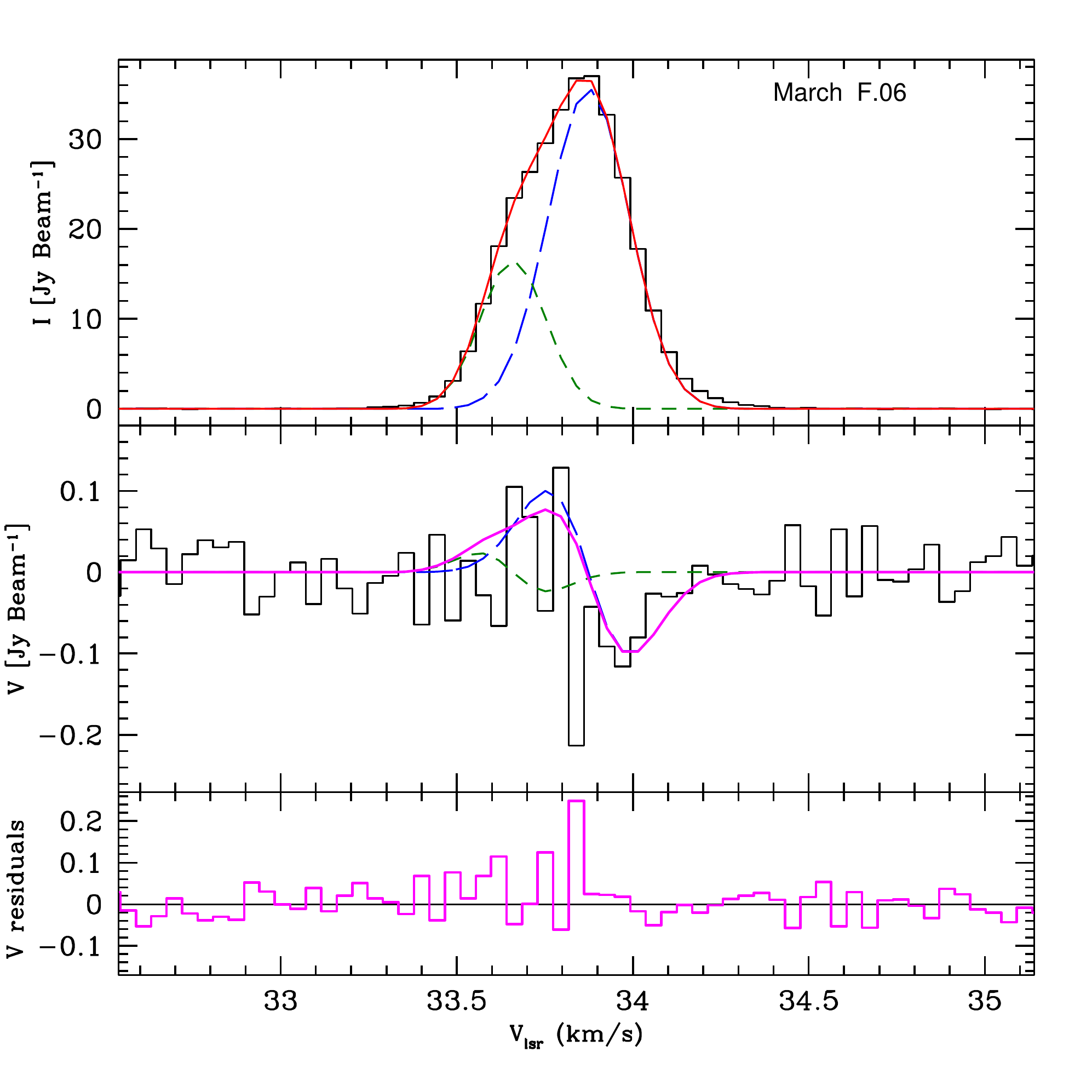}
   \includegraphics[width=.9\columnwidth]{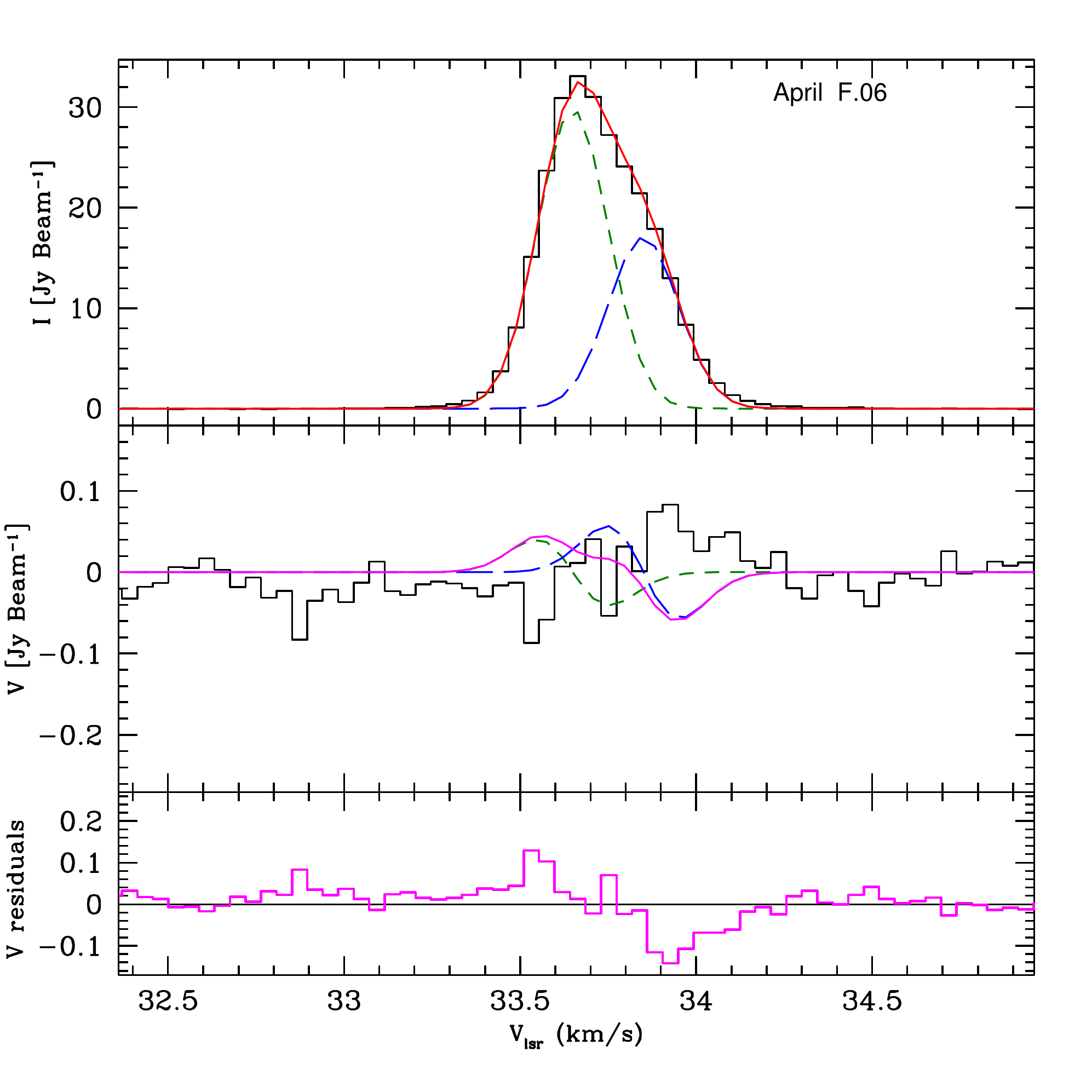}
   \caption{Fit to maser feature F.06 assuming the presence of two
     hyperfine components; left: March observation, right: April
     observation.
     Upper panels: total intensity I spectrum plotted as the black histogram.
     The solid red curve in the upper panel is the best fit using
     two Gaussian components, representing two maser hyperfine
     components separated by 0.2 km~s$^{-1}$ (see
     Sect.~\ref{sec:strength}). Component 1
     is the blue long-dashed line and component 2 is the green
     short-dashed line.
     Middle panels: circular
     polarisation V spectrum (black histogram). The solid magenta line
     is the sum of the derivatives of the two components of the fit
     from the upper panel. The blue long-dashed line is the derivative of component 1
     and the green
     short-dashed line is the derivative of component 2.
     Lower panels: we plot the residuals
     between the circular polarisation (middle panel, black histogram)
     and the sum of the derivatives (middle panel, solid magenta line).
        }      \label{Fig:hypercirular}
\end{figure*}

The strength of the magnetic field along the line of sight can be
obtained from the circular polarisation
\citep{Vlemmings2001,Vlemmings2006,Surcis2014H2O}.
Previous works showed that the circular polarisation fraction in
methanol masers is typically very weak (<1\%; e.g
\citealt{Surcis2015}), so circular polarisation can be observed more easily
in the brightest maser features. However,
as shown by \citet{Vlemmings2001, Vlemmings2002b}, an increase of the
noise, a narrowing and re-broadening, or a change in shape of the
maser line can also occur in hyperfine interactions or in 3D maser
propagation effect.  Therefore, as suggested by \citet{Surcis2015}, a
detection of circular polarisation should only be considered real if
it presents a V peak flux density at least
five times higher than the RMS.

In our observations, we found several of the above effects
that made it difficult to identify the
Zeeman effect. Therefore, we only propose a tentative detection of
circular polarisation for one maser feature (reported as F.06 in
Table~\ref{tab:maser_marchREF} and Table~\ref{tab:maser_aprilREF},
respectively) shown in Fig.~\ref{Fig:hypercirular}. For F.06 we
tentatively compute the magnetic field strength along the line of
sight.

The observed V spectrum is a sin-shaped function, corresponding to the
derivative I$^\prime$ of the total power spectrum I
\citep{Troland1982}. By fitting Gaussian components to the I spectrum,
and the corresponding derivative to the V spectrum, we can take
\begin{equation}
  \label{eq:PV}
V =aI+b \frac{\de I}{\de{\nu}}
,\end{equation}
where $a$ and
$b=zB\cos\theta$ are fit parameters, together with the Gaussian
components' intensity, centre velocity, and line width. $B$ is the
magnetic field strength, $\theta$ is the angle between the magnetic field
and the line of sight and $z$ is the Zeeman splitting factor for
CH$_3$OH. However, $z$ depends on the Land\'e $g$-factor, which was unknown
for the methanol maser molecules until recently. Therefore, all the previous
estimation of $B$ along the line of sight ($B_{los}$) were
affected by this uncertainty.  Recently a list of $z$ factor values has
been estimated for the 6.7~GHz methanol transition
$5_{15}$A$_{2}\leftarrow~6_{06}$A$_{1}$ and all its possible hyperfine
components \citep{Lankhaar2017}.

In Fig.~\ref{Fig:hypercirular}, we show the Stokes I profile (upper
panels, black histogram) for the maser feature F.06, in March (left
panel) and April (right panel).  We tried to
fit the line using two Gaussian components separated by
$\sim0.2$ km~s$^{-1}$, and we plot the best fit in red,  component 1
is the blue long-dashed line and  component 2 is the green
short-dashed line. We summarise the best fit parameters in
Table~\ref{tab:componenti}.

\begin{table*}
  \caption[]{Best-fit parameters for the two-component model for feature F.06.}
  \label{tab:componenti}
  \centering
  \begin{tabular}{ccccccc}
    \hline\hline
    Component         & \multicolumn{2}{c}{Intensity}         & \multicolumn{2}{c}{Centre Velocity} & \multicolumn{2}{c}{Velocity Linewidth}            \\
                      & \multicolumn{2}{c}{(Jy beam$^{-1}$)} & \multicolumn{2}{c}{(km~s$^{-1}$)}   & \multicolumn{2}{c}{(km~s$^{-1}$) }            \smallskip    \\
                      & March            & April           & March & April & March & April \\
    \hline
    Component 1 & 35.50               & 17.00              & 33.87     & 33.85     & 0.11     &0.10       \\
    Component 2 & 16.50                & 29.50              & 33.67     &33.65    & 0.09    & 0.10   \\
    \hline
  \end{tabular}
\end{table*}

It is possible to see that the I spectrum in April presents a reversed
profile with respect to March. A similar behaviour is also seen in the
circular polarisation spectra V, plotted as black histograms in the
middle panels of Fig.~\ref{Fig:hypercirular}: it presents two S-shape
profiles, one being the opposite of the other.

Following Lankhaar et al., a possible explanation of such I profiles
and opposite circular polarisation could be due to the presence of two
hyperfine components of the 6.7~GHz methanol transition
$5_{15}$A$_{2}\rightarrow~6_{06}$A$_{1}$: the F=$3\rightarrow~4$
(component 1) and the F=$4\rightarrow~5$ (component 2), also separated
by $\sim0.2$ km~s$^{-1}$ \citep{Lankhaar2016}. Under this hypothesis,
one hyperfine transition would be preferred over the other in one
epoch, and vice versa in the following epoch.  Therefore, we tried to fit
our V spectra using the sum of the derivatives of the two hyperfine
components (magenta line in the middle panels of
Fig.~\ref{Fig:hypercirular}). The derivatives of the single components
are also plotted: the derivative of component 1 is the blue
long-dashed line and the derivative of component 2 is the green
short-dashed line. The March V spectrum can be reproduced
(Fig.~\ref{Fig:hypercirular}, left middle panel) by using two
hyperfine components having a Zeeman coefficient $z$ of -1.135
Hz~mG$^{-1}$ for component 1 and -0.467 Hz~mG$^{-1}$ for component 2:
the resulting $B_{los}$ for both of them is $\sim5.7$ mG. In the right
middle panel we plot the V spectrum for April and the predicted
spectrum using a similar $B_{los}\sim5.5$ mG. Even taking into account
an increased noise in the April observations, the circular
polarisation spectrum is not consistent with the expected spectrum.

\begin{figure*}
   \centering
   \includegraphics[width=.9\columnwidth]{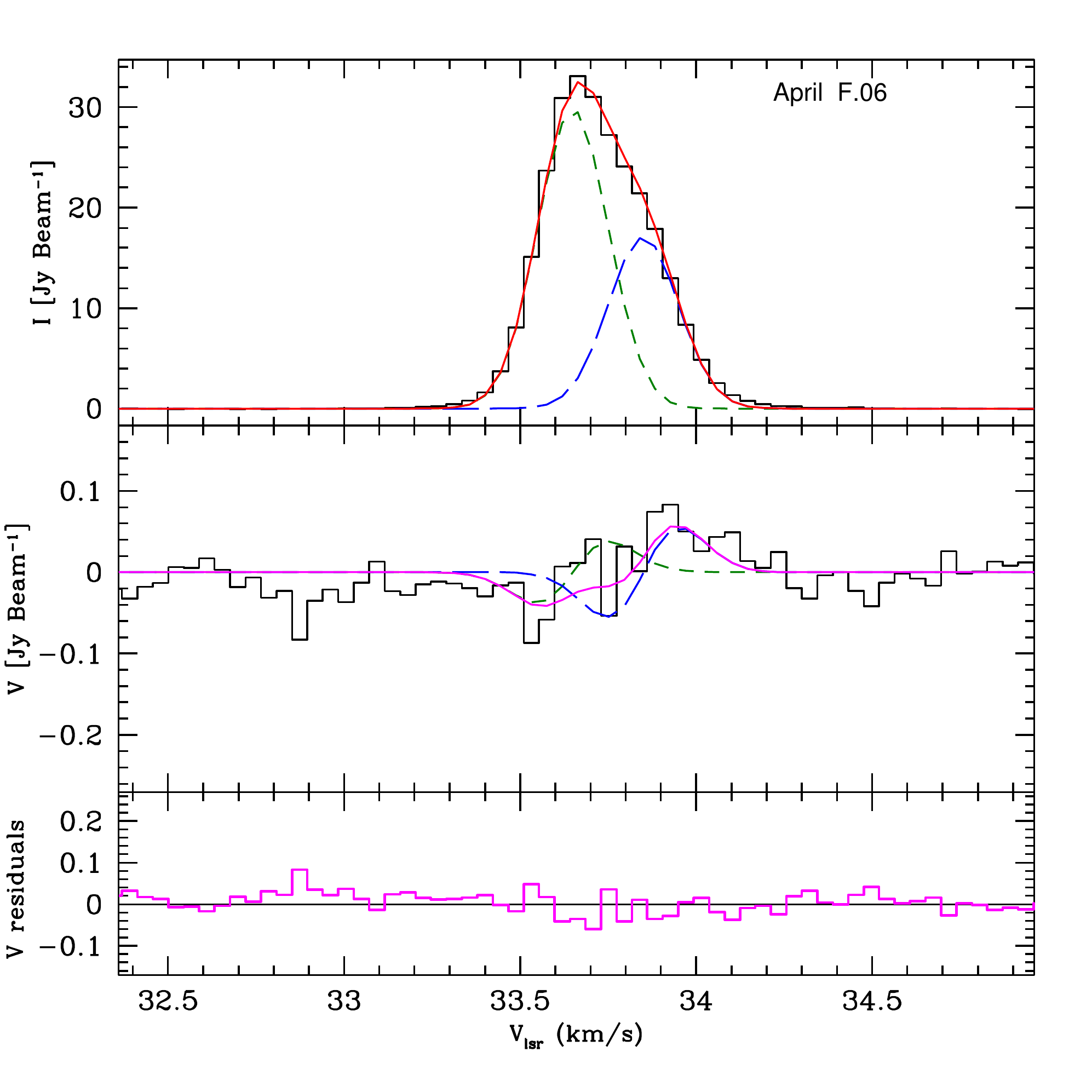}

   \caption{Fit to maser feature F.06 assuming the presence of two hyperfine components,
     but with a magnetic field that changes its sign between March and
     April. Since the March panel is identical to that in
     Fig.~\ref{Fig:hypercirular}, we only show the April panel. Panels
     and lines
     are as in Fig.~\ref{Fig:hypercirular}.}
              \label{Fig:reversed}
\end{figure*}

Another possible explanation could be to consider a change in the
magnetic field direction between March and April. We show this case in
Fig.~\ref{Fig:reversed}, where the two hyperfine components present the
inverse behaviour with respect to the previous case. The magnetic
field is still $|B_{los}|\sim5.5$ mG for both components. Similar variability of the
magnetic field has previously been observed and the origin of this change
can be intrinsic to the source as discussed by
\citet{Vlemmings2009}.

Finally, another possibility is to consider the two components as two
different and blended masers, originating in two places lying along the
same line of sight (e.g.  \citealt{Momjian2017}), and with varying
intensity. In this case we can only measure the average magnetic field
along the line of sight $B_{los} \sim 5.5$~mG in March, and
$B_{los} \sim -4.9$~mG in April. We report $B_{los}$ for this last
case in Tables~\ref{tab:maser_marchREF} and
\ref{tab:maser_aprilREF}.

However, it is impossible at this stage to discern which of these phenomena is
occurring. Higher resolution observations covering more epochs could
help to increase the signal-to-noise ratio and  better distinguish the
two components' contributions.

In all cases, the error on $B_{los}$ is approximately 30\% and was
estimated on the basis of the RMS noise in the line-free channels of
the V spectrum.  All the above values are comparable with
$B_{pos} \sim 11$~mG already obtained by \citet{Beuther2010}, from
dust continuum polarisation observations.  Considering
$B_{los}\sim5.5$ mG, we can obtain the total magnetic field strength
$B_{tot}\sim \sqrt{B_{pos}^2+B_{los}^2} \sim 12$~mG and since
$B_{tot}=B_{los}/\cos\theta$ we can estimate the angle between the
magnetic field and the line of sight $\theta=63^{\circ +8}_{
  -6}$. This angle is in agreement with our assumption that
$\theta > \theta_{crit}\sim 55^{\circ}$ (see Sec.~\ref{sec:results})
and it is consistent with the strength of our measured polarisation
fraction (e.g. \citealt{Surcis2015}).

\section{Conclusions}
\label{sec:conclusions}

In this paper, we present our investigation of the magnetic field morphology of the
well-known high-mass-star forming region IRAS 18089-1732. We analysed
a three-epoch MERLIN observation of the 6.7 GHz CH$_3$OH maser
generated in a region of a few au, close to the disc.

We identified nine masers in March and April and seven masers in July,
confirming almost all the maser features already seen, as well as some new
detections. The July observations were in dual circular polarisation only,
so we performed the linear polarisation analysis only on the masers
observed in the first two epochs, for which we measured the median
linear polarisation fraction (P$_l$) and the median linear
polarisation angle ($\chi$) across the spectrum. We identified two
groups of masers on the basis of two different velocities and $\chi$
values: a blue group spanning a velocity range from 30.0 to 36.4
km~s$^{-1}$, and a red group from 37.7 to 39.2 km~s$^{-1}$.

The two groups of masers showed ordered linear polarisation vectors,
and the orientation was preserved in both epochs. The blue group had a
weighted average angle of $\chi_{B,M} = -24^\circ \pm 8^\circ$ in
March and $\chi_{B,A} = -31^\circ \pm 12^\circ$ in April. The red
group had only one linear polarised emission in March with an angle
$\chi_{R,M} = -78^\circ \pm 5^\circ$, while in April the weighted
polarisation angle was $\chi_{R,A} = -70^\circ \pm 2^\circ$.

All three epochs were close to maxima if extrapolated from
\citet{Goedhart2009} but for two of the epochs the brightest feature
had a flux density much lower than predicted, suggesting irregular
periodicity or a change in magnitude or both. Our positions, more
accurate than those in \citet{Goedhart2009}, confirmed the separations
of features, and implied lower limits to light travel time that are in
some cases incompatible with the simplest interpretations of time
delays.

From the monitoring of \citet{Goedhart2009}, we noticed that all the
masers in the red group have variability with peaks occurring ahead of
that of the reference feature, while those in the blue group lag
behind. Since the two groups are separated in polarisation angles and
velocities, we deduce that the two groups of masers are emitted by two
different regions, one lying on the disc of the protostar and another
closer to the base of the molecular outflow. Therefore we suggest they are probing
two different magnetic field directions, with an orientation on the
plane of the sky of $\Phi^{disc} =+62^\circ\pm3^\circ$ and
$\Phi^{outflow} =+14^\circ\pm4^\circ$.

We showed that the small-scale magnetic field probed by the masers is
consistent with the large-scale magnetic field traced by the dust
\citep{Beuther2010}. Therefore we conclude that the large-scale field
component, even at the au scale of the masers, dominates over any
small-scale field fluctuations.

We proposed a tentative detection of circular polarisation for one of
the brightest features, F.06. The shape of the total power and of the
circular polarisation spectra appear to reverse between the March and
the April observations. This could be due to the splitting of two
hyperfine components, each one emitting preferentially in a different
epoch. Another possibility could be that the magnetic field reverses
its sign, as already suggested for another variable 6.7 GHz maser. Yet
another option could be that we observed two different masers,
originating in two places lying along the same line of sight. In all
the three cases we obtained a $|B_{los}|\sim5$ mG, comparable to
$B_{pos} \sim 11$~mG already obtained by \citet{Beuther2010} for dust.

\begin{acknowledgements}
  We thank the anonymous referee whose comments have contributed to
  improve the presentation of this paper. The research leading to these
  results has received funding from the European Research Council
  under the European Union's Seventh Framework Programme
  (FP/2007-2013) / ERC Grant Agreement n. 614264.
\end{acknowledgements}

\bibliographystyle{aa}
\bibliography{../bibliografia}

\end{document}